\documentclass[acmtog, authorversion]{acmart}
\copyrightyear{2023}
\acmYear{2023}
\setcopyright{rightsretained}\acmConference[SIGGRAPH '23 Conference Proceedings]{Special Interest Group on Computer Graphics and Interactive Techniques Conference Conference Proceedings}{August 6--10, 2023}{Los Angeles, CA, USA}
\acmBooktitle{Special Interest Group on Computer Graphics and Interactive Techniques Conference Conference Proceedings (SIGGRAPH '23 Conference Proceedings), August 6--10, 2023, Los Angeles, CA, USA}
\acmPrice{15.00}
\acmDOI{10.1145/3588432.3591487}
\acmISBN{979-8-4007-0159-7/23/08}

\AtBeginDocument{%
  \providecommand\BibTeX{{%
    \normalfont B\kern-0.5em{\scshape i\kern-0.25em b}\kern-0.8em\TeX}}}

\acmSubmissionID{166}

\begin{document}

%%
%% The "title" command has an optional parameter,
%% allowing the author to define a "short title" to be used in page headers.
\title{PMP: Learning to Physically Interact with Environments using Part-wise Motion Priors}

%%
%% The "author" command and its associated commands are used to define
%% the authors and their affiliations.
%% Of note is the shared affiliation of the first two authors, and the
%% "authornote" and "authornotemark" commands
%% used to denote shared contribution to the research.
\author{Jinseok Bae}
\affiliation{
  \institution{Seoul National University}
  \country{South Korea}
}
\email{capoo95@snu.ac.kr}

\author{Jungdam Won}
\affiliation{
  \institution{Seoul National University}
  \country{South Korea}
}
\email{nonaxis@gmail.com}

\author{Donggeun Lim}
\affiliation{
  \institution{Seoul National University}
  \country{South Korea}
}
\email{rms2836@snu.ac.kr}

\author{Cheol-Hui Min}
\affiliation{
  \institution{Seoul National University}
  \country{South Korea}
}
\email{mch5048@snu.ac.kr}

\author{Young Min Kim}
\affiliation{
  \institution{Seoul National University}
  \country{South Korea}
}
\email{youngmin.kim@snu.ac.kr}

%% a fresh cut. the current version is a bit dull, and needs more enthusiasm!
\begin{abstract}

We present a method to animate a character incorporating multiple part-wise motion priors (PMP). 
While previous works allow creating realistic articulated motions from reference data, the range of motion is largely limited by the available samples.
Especially for the interaction-rich scenarios, it is impractical to attempt acquiring every possible interacting motion, as the combination of physical parameters increases exponentially.
The proposed PMP allows us to assemble multiple part skills to animate a character, creating a diverse set of motions with different combinations of existing data.
In our pipeline, we can train an agent with a wide range of part-wise priors.
Therefore, each body part can obtain a kinematic insight of the style from the motion captures, or at the same time extract dynamics-related information from the additional part-specific simulation.
For example, we can first train a general interaction skill, \textit{e.g.} grasping, only for the dexterous part, and then combine the expert trajectories from the pre-trained agent with the kinematic priors of other limbs.
Eventually, our whole-body agent learns a novel physical interaction skill even with the absence of the object trajectories in the reference motion sequence.
% In the experiments, we demonstrate that our method can generate a diverse set of motions that perform rich interactions with the surroundings in the various kinds of complex scenarios. 
\end{abstract}

%%
%% The code below is generated by the tool at http://dl.acm.org/ccs.cfm.
%% Please copy and paste the code instead of the example below.
%%

% \ccsdesc[300]{Computer systems organization~Redundancy}
% \ccsdesc{Computer systems organization~Robotics}
% \ccsdesc[100]{Networks~Network reliability}

%%
%% Keywords. The author(s) should pick words that accurately describe
%% the work being presented. Separate the keywords with commas.

% \received{20 February 2007}
% \received[revised]{12 March 2009}
% \received[accepted]{5 June 2009}

%%
%% This command processes the author and affiliation and title
%% information and builds the first part of the formatted document.
\begin{teaserfigure}
\includegraphics[width=\textwidth]{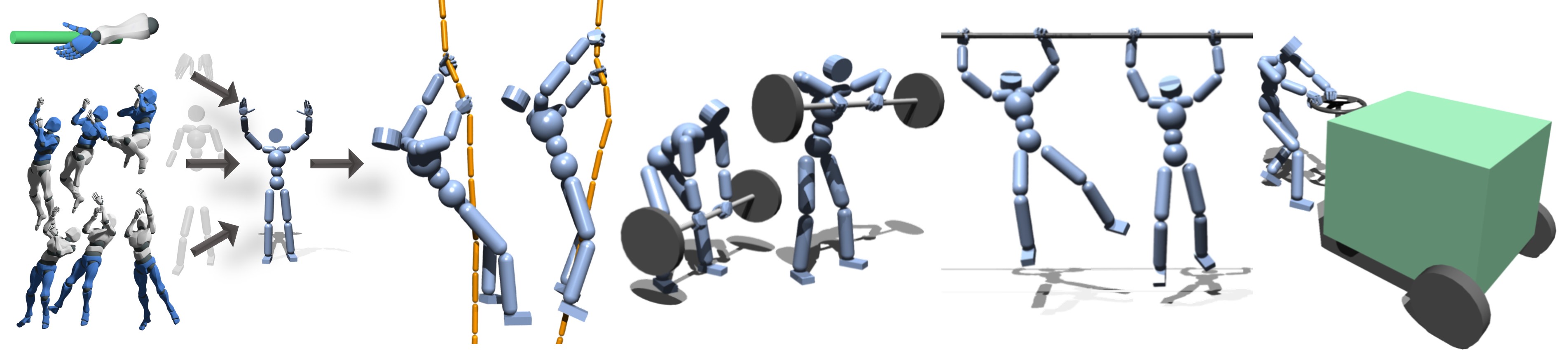}
\caption{Our method utilizes multiple part-wise motion priors to physically interact with the environments.}
\Description{figure description}
\end{teaserfigure}

\maketitle

\section{Introduction}
Motions for a virtual character are often created manually by an animator, or transferred from motion capture devices.
Both approaches are labor-intensive and it is not trivial to directly transfer the naturalness of novel characters or environments.
As a result, most consumer VR applications merely overlay isolated objects or people in the scene, unless the allowed motion sets are carefully designed beforehand. %in a pre-defined environment.
Such a lack of direct interaction fundamentally limits the extent of the immersive experience.

The physics-based animation utilizes a physics simulator to animate virtual characters in the environment with real-world dynamics.
However, the data-driven approaches are confined to creating motions that closely follow the joint states in the reference motion captures.
For example, when you only have normal locomotion and a staring idle motion separately in the database, the existing techniques cannot animate a character to run to the target while simultaneously staring at a specific location. 
Similarly, the range of human-object interactions cannot be fully covered by the samples available in the publicly available datasets.
Especially, the motion of hand joints is subtle with high degrees of freedom (DoFs) and suffers from frequent self-collision and severe occlusions, which makes it difficult to capture the hand motions.
A few prior works devised methods\cite{park2022handoccnet, zhang2021manipnet} to concurrently obtain the hand motion with the interacting object, but still the accurate capturing is challenging.
Nonetheless, the hand is crucial to accurately represent daily interaction.
Furthermore, the hand configuration can affect the overall balance and is tightly correlated to body movements, for example, in gymnastics and parkour.

In this work, we present a method to animate hand-equipped characters performing complex interactions with the environment.
We propose learning part-wise motion prior (PMP), and train an agent to combine a small number of expert motion sequences to perform a novel motion task. 
The physics simulator effectively balances the parts and deduces the unified policy of the entire body with rich repertoires compared to na\"ively mimicking the reference full-body motion.
By dissecting the problem into different parts, our pipeline can stably handle the imbalance prevalent in data, DoFs, and the range of motions among part segments.

Most importantly, our agent can obtain complex motor skill in a scalable way without requiring a large number of motion trajectories.
Strong prior on physical interaction can also come from another simulation for a part, and our method eventually assembles the demonstrations with the motion captures on the other side of the body.
Therefore with our approach, it is possible to convey an additional interaction prior to the virtual character's dexterous parts, \textit{i.e.} hands, from a single pre-training of the general interaction skill with a standalone training environment.
We showcase the generalizability of the proposed pipeline with various challenging scenarios with a humanoid exhibiting different part assignments. 
\section{Related Work}
We review recent studies that are closely related to our work such as data-driven methods for generating human-environment interaction, physics-based character controllers, and full-body character control including manipulation.

In computer animation research, data-driven approaches have been popularly used for generating motions with interaction. 
Researchers have tried to create complex interactions by combining several short pre-recorded motion clips including human-human or human-object interactions~\cite{lee2006motion, shum2008simulating, shum2010simulating, won2014generating}. 
Recently, approaches based on deep learning have shown promising results as large datasets which include human-scene interaction have become publicly available. 
Locomotion adapting to surrounding environments (e.g. uneven terrains) has been demonstrated, where the system gets either future trajectories~\cite{holden2017phase, zhang2018mode} or pose conditioning frames~\cite{harvey2020robust, tang2022real} as inputs. 
Deep neural networks that produce more general human-scene interactions (e.g. sitting on chairs or passing through doorways) can also be learned by using larger datasets that include those interactions~\cite{starke2019neural, hassan2021populating, wang2021synthesizing, wang2022towards}.
A few works have attempted to generate limited range of interaction motions along with the hands using whole-body motion captures~\cite{taheri2022goal,wu2022saga,ghosh2022imos, tendulkar2022flex}.
Although these methods produce plausible motions with interaction, they could suffer from artifacts such as foot sliding or penetration especially when the datasets are not well-prepared in advance.  

Physics-based methods have been investigated as an alternative to generating natural-looking motions because physical constraints ensure that the generated motions are physically plausible and they often generalize better to unseen scenarios.  
Especially for the motion tracking task given reference motions, a number of works showed that deep reinforcement learning (DRL) can provide near-perfect results~\cite{2018-TOG-deepMimic, bergamin2019drecon, park2019learning, won2020scalable, peng2021amp, fussell2021supertrack}.
On the other hand, developing physics-based controllers for reproducing complex physical interactions still has been considered as a challenging problem, and only a few studies have been demonstrated for human-human~\cite{won2021control, bansal2017emergent}, human-object~\cite{liu2018learning, merel2020catch, yang2022learning} interaction. 
The most closely relevant studies to our method are~\cite{liu2018learning, merel2020catch}, which developed controllers of full-body characters with dexterous hands. 
Merel et al.~\shortcite{merel2020catch} proposed a vision-based controller that can catch and carry objects with hands. 
Although the framework is general enough to be applied to many other tasks, the motion quality is less comparable to the examples demonstrated in other motion tracking controllers, more importantly, it requires paired input motions where human actors and objects are recorded simultaneously, which are often unavailable. 
Liu et al.~\shortcite{liu2018learning} used two separate controllers for the body and arms, which are trained in the different pipelines, and the movement of the target object are reconstructed via trajectory optimization.
This framework, however,  does not assume the existence of paired motions, and requires curating a large amount of task-specific heuristics.
Our method combines the benefits of both frameworks, where it can be easily applied for various applications and it uses actor motions only.

The key idea of our method is providing extra freedom for controllers to reassemble given reference motions, for which we develop part-wise motion priors.
A few studies have also exhibited similar idea on part-wise motion assembly.
Hecker et al.~\shortcite{hecker2008real} proposed a retargeting algorithm for highly varied user-created characters, where morphology-agnostic semantics of animations recorded during animation authoring are transferred to unknown target characters via inverse kinematics.
Jang et al.~\shortcite{jang2022motion} developed a method called \textit{Motion Puzzle} that performs part-wise style transfer of a source motion to a target motion by learning part-wise style networks and a graph convolutional network to extract motion features.
These are kinematic methods, so the output motions could have similar artifacts mentioned above.
Lee et al.~\shortcite{lee2022learning} developed physics-based controllers for new characters created by reassembling body parts of various characters, where the controllers are jointly optimized via supervised (part assembly) and reinforcement learning (dynamic control). 
Similarly, our method also provides a physically plausible way for part-wise assembly but our formulation is more flexible in a sense that it allows assembly of non-periodic motions with finer control rate.
\section{Kinematic Motion Prior}\label{sec:kinematic_motion_prior}

The kinematic motion prior comes from the expert trajectories that are provided by motion capture data.
We first revisit the core idea of AMP~\cite{peng2021amp} (Sec.~\ref{sec:amp}), which successfully utilizes the kinematic prior of motion in combination with RL to perform natural motion within a physics simulator. 
Then we explain our proposed part-wise assembly which efficiently excavates meaningful motion skills from multiple sources of motion capture data (Sec.~\ref{sec:part-wise}).

\subsection{Adversarial Motion Prior (AMP)}\label{sec:amp}

Instead of the tedious process of the reward design for naturalness, AMP~\cite{peng2021amp} utilizes the reference data to infer the reward signal for natural motion.
The total reward $r$ is a weighted sum of the task reward $r^g$ and the style reward $r^s$, which is
\begin{equation}
  r=w^g r^g + w^s r^s.
\end{equation}\label{eq:amp_total_reward}
The task reward is a typical reward of RL, representing how well the agent has achieved a given goal.
The style reward adapts the idea of Generative Adversarial Imitation Learning (GAIL)~\cite{ho2016generative} and performs inverse RL to match the kinematic style of the reference motions in the dataset.
Specifically, the style reward is calculated from the output of the discriminator $D_\phi$ as follows:
\begin{equation}
  r^s=-c\cdot\log{(1-D_\phi(o,o'))},
\end{equation}\label{eq:amp_style_reward}
where the tuple of $o$, and $o'$ corresponds to the observations in the neighboring timesteps, and $c$ is a scaling coefficient.
The discriminator is trained to mainly minimize the label prediction error between the generated motion and the demo trajectory:
\begin{equation}
    \label{eq:amp_pred_loss}
  \underset{\phi}{\mathrm{argmin}}~{\mathbb{E}_{\mathcal{M}}[-\log{(D_\phi{(o, o'))}}] + \mathbb{E}_{\pi_\theta}[-\log{(1 - D_\phi{(o, o'))}}]},
\end{equation}
where $\mathcal{M}$ is a set of demonstrations, and $\pi_\theta$ is a policy of the agent.

The success of AMP is confined within the extent of available observations $o$. 
The observations should account for the underlying DoFs of the agent performing the task.
DoFs excluded in the observation are merely optimized for the task reward, and cannot be guided for natural motion.
Consequently, the agent is less likely to achieve good performance in the task unless the dataset contains the motion directly related to the scenario.
However, it is intractable for the off-the-shelf motion clips to cover the numerous real-world interaction scenarios.

\subsection{Part-wise Motion Priors (PMP)}\label{sec:part-wise}
We propose obtaining strong part-wise motion priors that can be mixed into sophisticated skills for a wide range of interactions.
A na\"ive strategy to avoid data insufficiency is generating an augmented motion database in advance via kinematics approach~\cite{zhao2020bayesian, petrovich2021action} and using it as a source to train the discriminator.
However, the possible combinations of motion mixtures are infinite and they are not guaranteed to be physically plausible.
We instead leverage the existing motion capture data for different parts and assemble them efficiently. 
Our method guides each part to refer the part-specialized prior, then allows the agent to explore and dynamically select the holistic skill that best fits the scenario during the training phase.
This combinatorial approach is also a practical choice when the movements of all the joints are not stored in the given reference motion.

Our method partitions the full list of joint $J$ into $K$ sets 
\begin{equation}
  J = \bigcup_{k=1}^K J_k.
\end{equation}\label{eq:joint_union}
For example, if we segment the body into the upper body (without hands), lower body, and two hands, then $K=3$.
Note that there is no restriction on the choice of sets, and the joints in the same set are not required to be spatially connected to each other in the skeleton tree.
The core idea is to assign dedicated discriminators $D_{\phi_k}$ for the different sets of joints, instead of training a single discriminator $D_\phi$ for the whole body motion.
The optimization objective in Eq.~(\ref{eq:amp_pred_loss}) is modified as
\begin{equation}
    \label{eq:ours_pred_loss}
  \underset{\{{\phi_k}\}_K}{\mathrm{argmin}}~\sum_{k=1}^K\left\{\mathbb{E}_{\mathcal{M}_k}[-\log{(D_{\phi_k})}] + \mathbb{E}_{\pi_\theta}[-\log{(1 - D_{\phi_k}})]\right\},
\end{equation}
where $\mathcal{M}_k$ denotes the motion dataset for the $k$-th body part $J_k$, and $D_{\phi_k}=D_{\phi_k}(o_k,{o_k}')$ denotes the prediction probability of the $k$-th discriminator given the tuples of the partial observation $(o_k, {o_k}')$ from either $\mathcal{M}_k$ or $\pi_\theta$.

Although the discriminators are independently optimized, all the DoFs of the agents are simultaneously controlled with a unified policy as in Eq~(\ref{eq:ours_pred_loss}).
The style reward $r^s$ aggregates the $K$ terms as
\begin{equation}
    \label{eq:ours_style_reward1}
    r^s = c\cdot\prod_{k=1}^K r^s_k,
\end{equation}
where $r^s_k=-\log{(1-D_{\phi_k}(o_k,{o_k}'))}$.
We empirically found that the agent can coordinate different parts with the assistance of the physics simulation, and generate natural whole-body motion for a wide variety of scenarios.
However, an extremely small reward signal in one of the parts can diminish the entire style reward after the multiplication, especially when $K$ is large.
We prevent the reward from vanishing with a demo blend technique (Sec.~\ref{sec:training}), which serves a critical role to stabilize training.

\section{Interaction Prior}
\label{sec:interaction_prior}

Although the kinematic states in the motion capture provide essential information for the control, they don't transfer the knowledge on the human-object interaction unless the trajectories of paired objects are included in the data.
Our method can wisely bypass the addressed problem by incorporating an interaction prior obtained from a training of hand-only agent in a minimal environment.
Recall that our PMP framework allows each motion prior to have a different feature set, which means the dexterous hand can be benefited from completely different types of data other than mocaps such as trajectories from the simulation.
Interaction prior enhances the training efficiency, similar to the pre-training strategy for reusable effective skills in~\cite{hasenclever2020comic, won2022physics, peng2022ase}.
Section~\ref{sec:interaction_gym} introduces a training environment for a hand to obtain the interaction prior, the grasping skill, and Section~\ref{sec:interaction_state} contains the state definition to represent general grasping.
Then Sec.~\ref{sec:integration} describes how we can combine the interaction prior with the part-wise kinematic priors from Sec.~\ref{sec:part-wise}.

\begin{figure}[t]
  \centering
  \includegraphics[width=\linewidth]{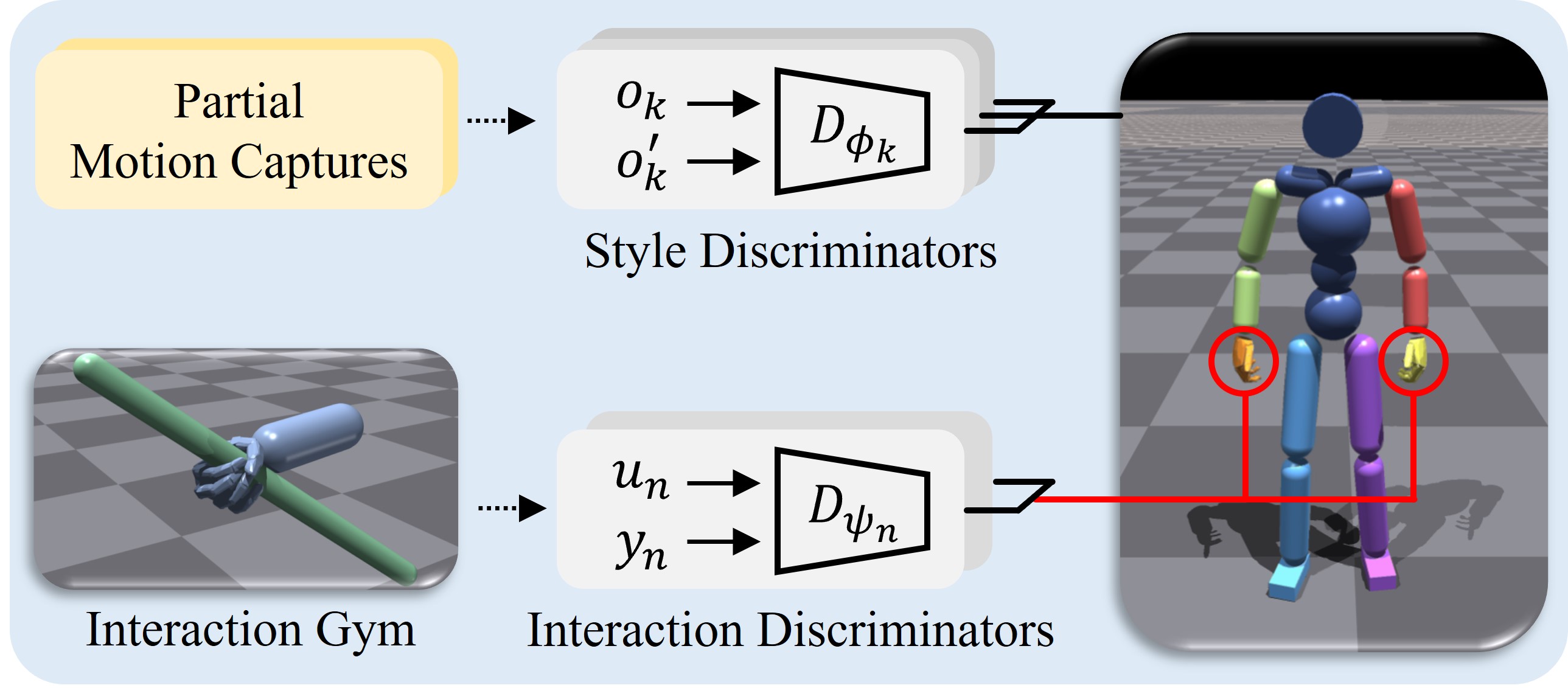}
    \caption{Visualization of the pipeline in our system. Kinematic style discriminators $\{D_{\phi_k}\}_K$ are trained with part-wise motion captures and interaction discriminators $\{D_{\psi_n}\}_N$ are trained with demo trajectories from the pretrained interaction gym. Note partial observations $\{o_k\}_K,\{u_n\}_N$ and hand actions $\{y_n\}_N$ are subsets of state $s$ and action $a$ of the whole-body agent.}
  \label{fig:pipeline}
\end{figure}  

\subsection{Interaction Gym}\label{sec:interaction_gym}
The interaction gym collects the state-action pairs of hand joints in a simulator, which can provide hand-specific supervision signals for various interaction-rich tasks.
The simulator set-up is composed of a hand model and a representative target object, as shown in Fig.~\ref{fig:pipeline}.
We use a cylindrical rod for the grasping target, and the grasping skill is utilizable to more general objects as we discuss in the results.
Since the controllable DoF of the hand is significantly smaller than that of the whole-body agent, we apply an RL algorithm with manually designed rewards to train natural and physically stable grasping. 
Out of many possible grasping styles, we enforce the hand to hold a target object with even contact to it.
We further encourage stable grasping by applying an arbitrary force and torque to the target object throughout the episode.

\subsection{Interaction State}\label{sec:interaction_state}
The state $s^i$ is the input to the policy to output action $a^i$, and 
it is crucial to have a proper part representation to learn a powerful and general interaction skill.
The interaction state of the hand $s^i$ contains both the proprioceptive information of the actuators and the physical relationship to the target rod
\begin{equation}
    \label{eq:interaction_state}
  s^i=\{q_{h},~\dot{q}_{h},~\bar{p}_{e},~\bar{p}_{r},~c_{e},~\langle d_{h}\cdot d_{c}\rangle\}.
\end{equation}
The first three terms contain the status of the hand.
We use the position $q_h$ and velocity  $\dot{q}_h$ of the hand joints, and additionally provide 3D Cartesian positions of the end-effector $\bar{p}_{e}$ for the index, middle, ring and pinky fingers in the local coordinate of the wrist.
Then, we represent the pose of the target rod $\bar{p}_{r}$ as a 6-dimensional vector, composed of the two end points of the rod in the wrist's local coordinate, similar to $\bar{p}_{e}$.
The last two terms sense the current state of the interaction.
$c_{e}$ denotes the binary contact markers for individual fingertips.
Finally, we find the cosine similarity between the vector pointing the inside of the hand from the fingers $d_h$ and the direction of the generated contact forces $d_c$ for every rigid body comprising each hand.

\subsection{Integration with Kinematic Prior}\label{sec:integration}

After pre-training in a minimal gym, we need to transfer the learned grasping skill to the whole-body agent to successfully perform an interaction-rich task.
However, the pretrained policy is not yet generalized by itself as it is trained only on the specific cylindrical rod.
We utilize the adversarial imitation learning such that the agent can adapt the acquired skill to downstream tasks.
Our part-wise motion priors in Sec.~\ref{sec:kinematic_motion_prior} nicely fits to the integration purpose.
The expert state-action pairs ($s^i, a^i$) from the gym serve as another set of partial demonstrations, and we can easily blend the interaction prior for the two hands with the corresponding discriminators $D_{\psi_n}$, where $n\in H, H=\{\text{right}, \text{left}\}$  as shown in Fig.~\ref{fig:pipeline}.

The kinematic hand prior is also necessary to generate natural preceding motions before the actual grasping.
While the interaction prior serves a critical role for the grasping, the demo sequences of the gym only contains motion that is directly related to the grasping action.
To make the hands naturally approach the object and initiate the grasp, we embed the interaction reward $r^i_n$ within the style reward term in Eq.~\ref{eq:ours_style_reward1} as
\begin{equation}
    r^i_n=-\log{(1-D_{\psi_n}(u_n,y_n))}, \mbox{ and }
\end{equation}\label{eq:interaction_reward}
\begin{equation}
    r^s = c\cdot\prod_{n\in H}\{(1-\sigma_n)r^s_n + \sigma_n r^i_n\}\cdot\prod_{k\notin{H}} r^s_k.
\end{equation}\label{eq:ours_style_reward2}
Here, $u_n$ and $y_n$ are subset of $s^i$ and $a^i$, respectively.
$\sigma_n$ is a coefficient from the Gaussian kernel $\Phi(u_n)$ based on the Euclidean distance such that the likelihood of the interaction is normalized into the scale of $[0, 1]$.
In this way, an agent receives more feedback from $D_\psi$ when its hand approaches to the target.
The reward formulation faithfully reflects both the style prior and interaction prior.
Further details including the implementation of $\Phi(u_n)$ are explained in the supplementary material.
\section{Training Techniques}\label{sec:training}

We use reinforcement learning to train both the unified policy of the whole-body integration (Sec.~\ref{sec:part-wise} and Sec.~\ref{sec:integration}) and the hand-only policy in the interaction gym (Sec.~\ref{sec:interaction_gym}).
While most of the training settings are inspired by the previous works on physics-based animation~\cite{2018-TOG-deepMimic, peng2021amp}, we make two important modifications to adapt to our part-wise setting.

\subsection {State Initialization}
The state initialization within a reference motion is an effective technique to allow a character to imitate a motion clip under a RL framework~\cite{2018-TOG-deepMimic}.
However, our agent refers to different sets of motion clips for individual part segments, and does not have an access to a reliable whole-body pose for the initialization.
We independently sample reference poses for parts without attempting to seek an optimal alignment.
We empirically found that the agent eventually coordinates between parts and discovers a natural whole-body motion for various tasks.

\begin{figure*}[t]
  \centering
  \includegraphics[width=\textwidth]{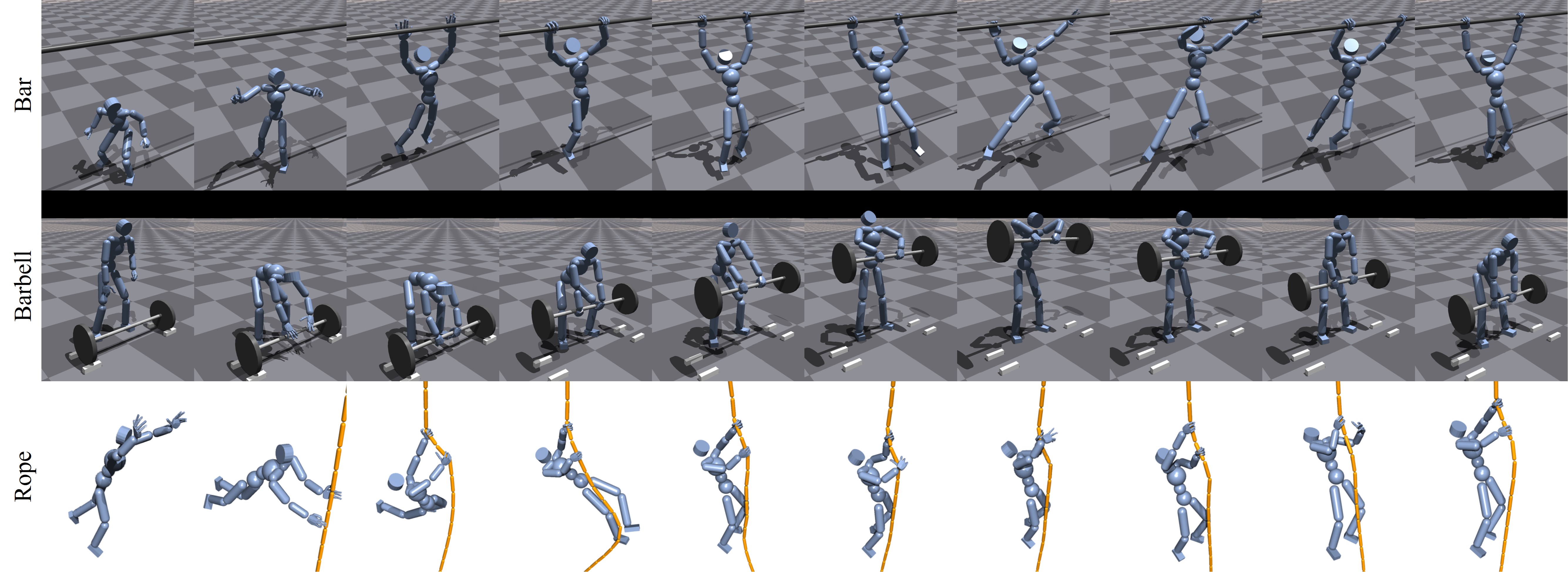}
  \caption{Results of training with PMP in the selected scenarios: \textit{Bar Hanging}, \textit{Barbell Lifting}, \textit{Rope Climbing}.}
  \label{fig:interaction_scenes}
\end{figure*}

\subsection{Demo Blend Technique}\label{sec:demo_blend}

Our style reward is the product of part-wise rewards $r^s_k$ from multiple parts.
Intention of using multiplication rather than summation is to ensure that priors from all segmented parts contribute to the global motion. 
However, the reward signal may vanish if at least one of the parts fails to imitate the reference motion.
It happens more frequently as the number of part segments increases, and halts the training process.
To address this issue, our \textit{demo blend technique}  randomly mixes experienced trajectories for individual parts.
Each substitution for a part happens independently in a Bernoulli distribution with a parameter $\lambda_d$.
The demo blend technique is especially useful in training an agent with a large number of decomposed parts.
In our experiment, we apply demo blending with the probability of  $\lambda_d=0.1$ for the scenario where the number of segmented parts is higher than two.
The efficacy is demonstrated with the empirical results in Sec.~\ref{sec:experiments}.

\section{Experiments}\label{sec:experiments}

The policy $\pi_\theta$ of our agent outputs an action with the frequency of 30 Hz as a proportional-derivative (PD) target.
We apply Proximal Policy Optimization (PPO)~\cite{schulman2017proximal}, similar to other works on the physics-based animation.
For the style reward, we collect the reference motions for each scenario from Mixamo~\cite{mixamo} and retarget them to our whole-body agent.
We set weights for the task reward $r^g$ and style reward $r^s$ as $w^g=0.5$ and $w^s=0.5$.

Our whole-body agent is modified from the humanoid of Deepmimic~\cite{2018-TOG-deepMimic}.
We replace the sphere-shaped hands of the original humanoid with the hand from Modular Prosthetic Limb (MPL)~\cite{kumar2015mujoco}.
The total actuated DoF is 54, where each hand has 3 DoF for its wrist and 10 DoF for its fingers.

Our agents are trained with the physics simulation of the \textit{Isaac Gym}~\cite{makoviychuk2021isaac} framework.
Its GPU acceleration can simultaneously train the agents under 4096 environments.
However, the efficiency of the highly parallel environment comes at the cost of losing detailed contact information, which is crucial in most of the contact-rich tasks.
We observed that our pre-traing handles the nuances of subtle contacts for the whole-body agents such that the agent can interact with dexterous hands without additional information such as mesh-level contact positions.

\subsection{Tasks}

To demonstrate the general advantage of our part-wise motion prior, we show the motion quality on seven different tasks as shown in Table~\ref{tab:part segments}.
The first three examples, \textit{Upstair Carrying}, \textit{Sight Locomotion}, and \textit{Walking Styles}, demonstrate that PMP can efficiently assemble the part-wise motions and generate a plausible new motion to solve a novel task.
The four other scenarios, \textit{Cart Pulling}, \textit{Bar Hanging}, \textit{Barbell Lifting}, \textit{Rope Climbing}, utilize the interaction prior to solve the complex interaction-rich tasks. 
When we incorporate the pre-trained interaction prior, we expand the state vector with the interaction state (Sec.~\ref{sec:interaction_state}) calculated from both hands.

Table~\ref{tab:part segments} also contains the different part segments used for the presented scenarios. 
\textit{Upper} includes abdomen, neck, shoulders, elbows, and wrists whereas \textit{Lower} includes hips, knees, and ankles.
Additionally, \textit{Trunk} denotes only abdomen and neck joints, while \textit{Limbs} represents shoulders, elbows, hips, knees, and ankles.
For interaction scenarios, we group wrists with \textit{hands} such that the motion of wrists can refer to the interaction prior.
We briefly describe the individual tasks in the following. The full rewards and states are available in the supplementary material.

\begin{small}
\begin{table}[h]
  \caption{Part segments for each scenario.}
  \label{tab:part segments}
  \begin{tabular}{c|cl}
    \toprule
    Experiments & Part Segments ($K$)\\
    \midrule
    \textit{Upstair Carrying}    & Upper - Lower (2) \\
    \textit{Sight Locomotion}     & Trunk - Limbs (2) \\
    \textit{Walking Styles}    & Trunk - R/L Arms - R/L Legs - R/L Hands (7) \\
    \textit{Cart Pulling}      & Body - Hands (2) \\
    \textit{Bar Hanging}      & Upper - Lower - Hands (3) \\
    \textit{Barbell Lifting}   & Upper - Lower - Hands (3) \\
    \textit{Rope Climbing}     & Upper - R/L Hands (3) \\
  \bottomrule
\end{tabular}
\end{table}
\end{small}

\begin{small}
\begin{table}
  \caption{Comparison of normalized average returns in PMP with interaction prior (PMP), without interaction prior (PMP (no IP)), and without PMP (no PMP). The values in the parenthesis refers to standard deviation.}
  \label{tab:comparison}
  \begin{tabular}{cccc}
    \toprule
    Scenario & PMP & PMP (no IP) & no PMP\\
    \midrule
    \textit{Upstair Carrying}     & -                     & \textbf{0.51} (0.047) & 0.50 (0.022) \\
    \textit{Sight Locomotion}     & -                     & \textbf{0.45} (0.072) & 0.39 (0.055) \\
    \textit{Cart Pulling} (Plain) & \textbf{0.65} (0.019) & 0.63 (0.021)     & 0.59 (0.012) \\
    \textit{Cart Pulling} (Random)& \textbf{0.65} (0.041) &  9.2e-5 (5.9e-5) &0.0066 (0.0035) \\
    \textit{Cart Pulling} (Curved)& \textbf{0.69} (0.010) &  0.63 (0.0060)   & 0.59 (0.020)\\
    \textit{Bar Hanging}          & \textbf{0.63} (0.25)  & 0.33 (0.16)      & 0.27 (0.19) \\
    \textit{Barbell Lifting}      & \textbf{0.55} (0.27)  & 0.052 (0.014)   & 0.022 (0.019)\\
    \textit{Rope Climbing}        & \textbf{0.63} (0.062) & 0.38 (0.13)     & 0.31 (0.17)  \\
    \textit{Rope Climbing} (Low $f$)& \textbf{0.26} (0.21)   & 0.22 (0.20)   & 0.063 (0.074) \\
  \bottomrule
\end{tabular}
\end{table}
\end{small}

\paragraph{Upstair Carrying}
The task is to walk down the ground plane to reach a stairway and then walk up to the highest stair, while carrying a ball in arms throughout the sequence.
Our pipeline can fulfill the task by referring to motions from the \textit{locomotion}, \textit{walking up the stair} for the lower body, and \textit{carrying idle} (carrying a ball while standing) for the upper body.
The states and task rewards are similar to the setting in \textit{target location} task in~\cite{peng2021amp}, but we augment the state vector with the $8\times8$ height map to observe the stair.
We penalize the agent dropping the ball by early termination~\cite{2018-TOG-deepMimic} as an indirect reward for carrying.

\paragraph{Sight Locomotion}
In this task, the agent tries to reach the target location (root goal) while looking at a sight goal.
The root goal and sight goal are independently respawned.
Available motion captures are the \textit{locomotion} and \textit{looking around}, which we imitate with the style reward for the limbs and the trunk, respectively.
The state and rewards are again the same as the \textit{target location} task, with an additional reward for the sight tracking.
The sight tracking reward measures the deviation of the current sight from the sight goal.

\paragraph{Walking Styles}
This scenario is designed to demonstrate that the part-wise prior can augment the motion trajectories in novel styles.
We deliberately used a large number of part segments - seven parts.
In addition to the dedicated discriminators for individual parts, we allocate one more discriminator, which enforces a set of specified joints to follow the demo trajectories.
We choose the joints of the shoulders and hips to receive signal from the additional discriminator, as those joints are critical for the temporal correlations of the limb movements.
We train the agent with three different styles of walking separately, namely \textit{normal}, \textit{soldier}, and \textit{hopping}.
The task reward is similar to the \textit{target heading} task in~\cite{peng2021amp}.

\paragraph{Cart Pulling}
The task is to bring a 30 kg wagon cart to the goal location by pulling its handle.
There are four different handles, namely a plain bar, a randomly-oriented plain bar, a thick bar, and a curved handle, where all four assets are unseen during the pre-training procedure.
To highlight the applicability, we additionally train the agent to pull the cart in two different grip orientations on the curved handle.
Here, we use \textit{pulling} motions as a reference motion.
In addition to the conventional state vector, we concatenate the cart states and the target position.
The rewards encourage the hands to approach the grip, and the cart to reach the goal location.

\paragraph{Bar Hanging}
The agent first jumps from the ground to hang on the 2 m-high bar with two hands, then horizontally shifts for 60 cm to the right or left direction along the bar according to the target grasping sites.
The reference motions are \textit{jumping to hanging} and \textit{horizontal hopping}.
The reward enforces the hands to reach their target grasping sites.
To accelerate the training, we terminate an episode  if the feet of the agent is still in contact with the ground after a few seconds from the start.

\begin{figure}
  \centering
  \includegraphics[width=\linewidth]{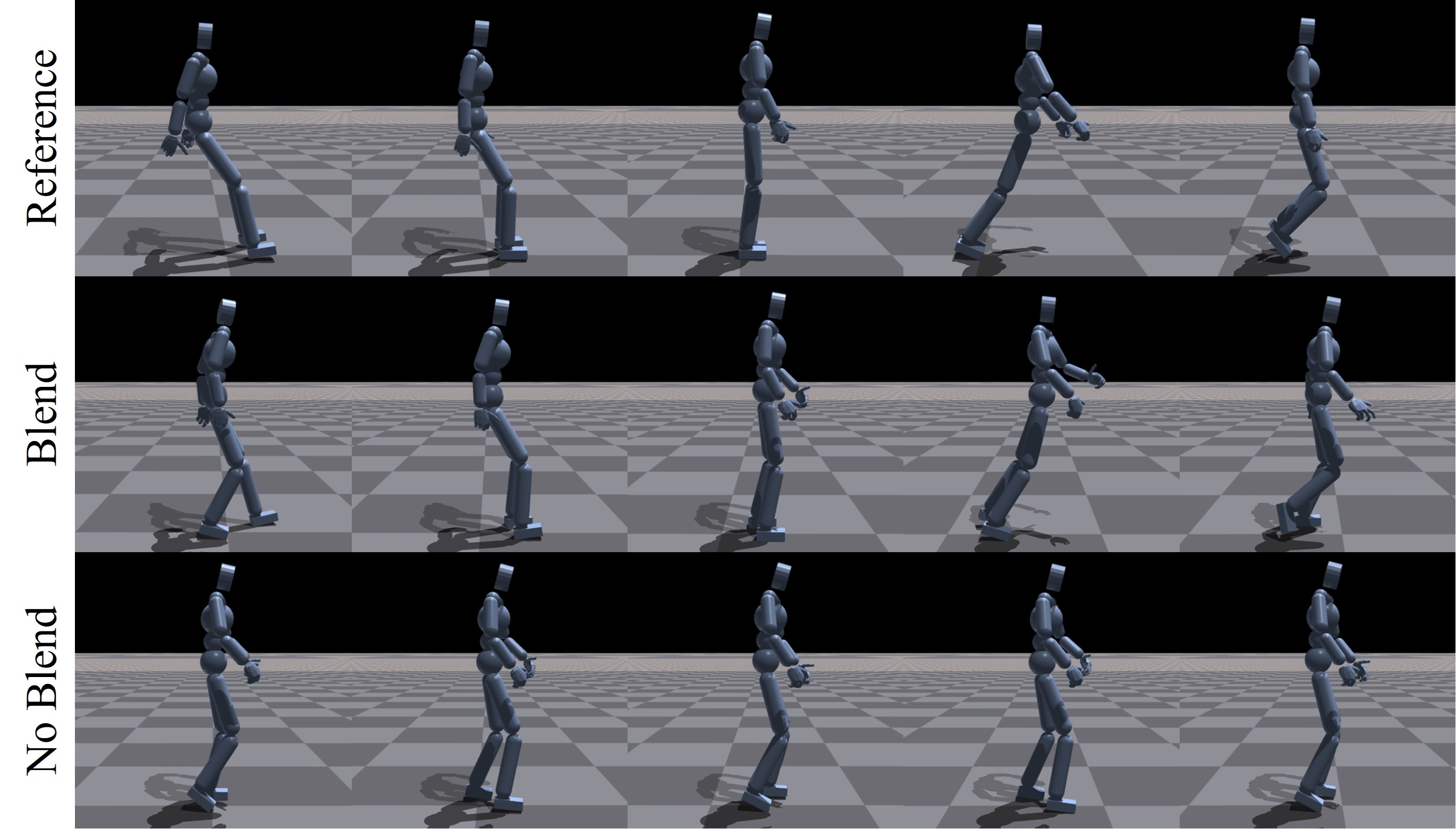}
  \caption{Results of walking imitation for \textit{hopping} style with PMP. Without demo blend technique, agent fails to imitate details of hopping motion.}
  \label{fig:demo_blend_result}
\end{figure} 

\begin{figure}[t]
  \includegraphics[width=\linewidth]{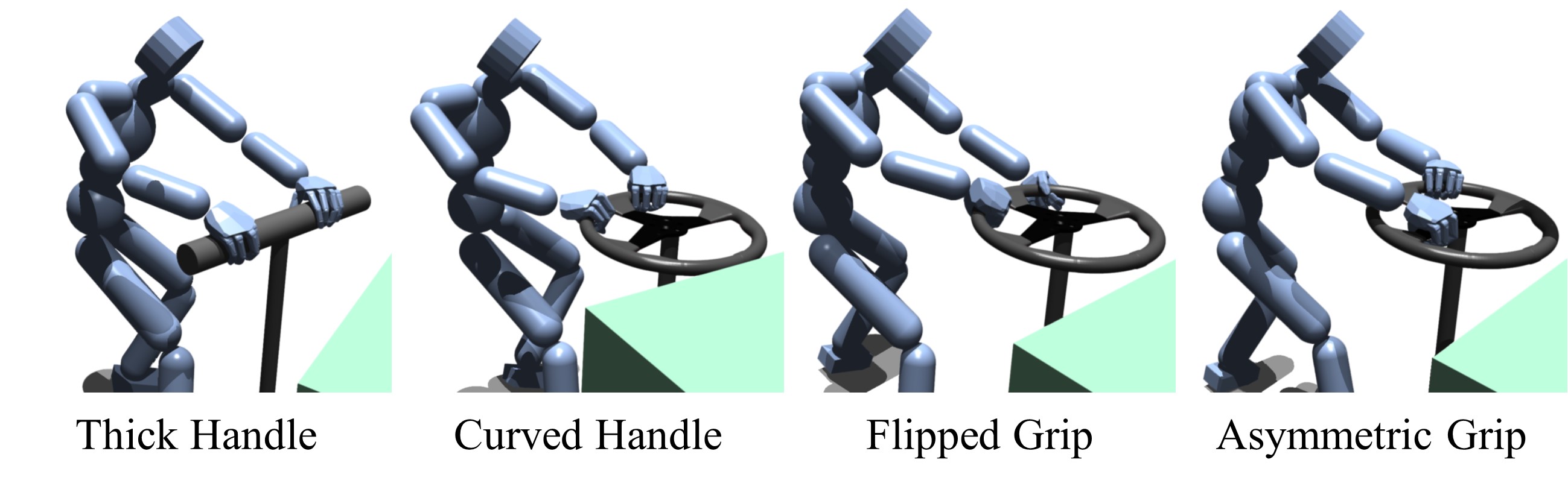}
  \caption{Snapshots on the sub-scenarios of \textit{Cart Pulling}. Agent adaptively uses interaction prior to grasp various shapes and grips.}
  \label{fig:pullcart_shapes}
\end{figure}

\paragraph{Barbell Lifting}
This example performs an exercise of `sumo pull' with a 10 kg barbell.
An agent first pulls up the barbell, then brings it at the target height.
The target height starts at 1.25 m, then alternates between 0.25 m and 1.25 m  throughout the episode.
The reference motion is  \textit{sumo pulling} for both the upper and lower body segments.
The state and rewards are similar to the  \textit{Cart Pulling}.

\paragraph{Rope Climbing}
The agent first approaches a rope from the air and grasps it.
The agent then climbs up the rope, which is induced by changing the target grip positions with the input command.
We intentionally ignore the lower body control, and solely rely on hands to climb the rope.
The reference motion for the upper body is \textit{rope climbing} motion.
Because the pre-grasp period is extremely short, only \textit{idle} motion is utilized for each hand.
The state vector is augmented with the rope state and the distances between fingers to the target grasping site.

\begin{figure*}
  \includegraphics[width=\textwidth]{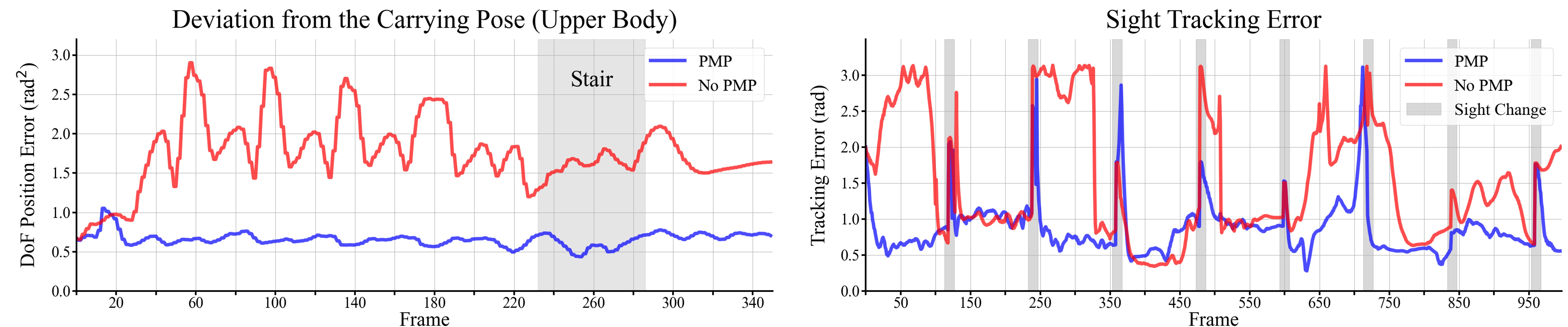}
  \caption{Evaluation on the \textit{Upstair Carrying} (left) and \textit{Sight Locomotion} (right). Note the position of the sight goal in \textit{Sight Locomotion} changes for every 120 frames.}
  \label{fig:locosight_staircarry_graph}
\end{figure*}

\begin{figure*}[h]
  \includegraphics[width=\textwidth]{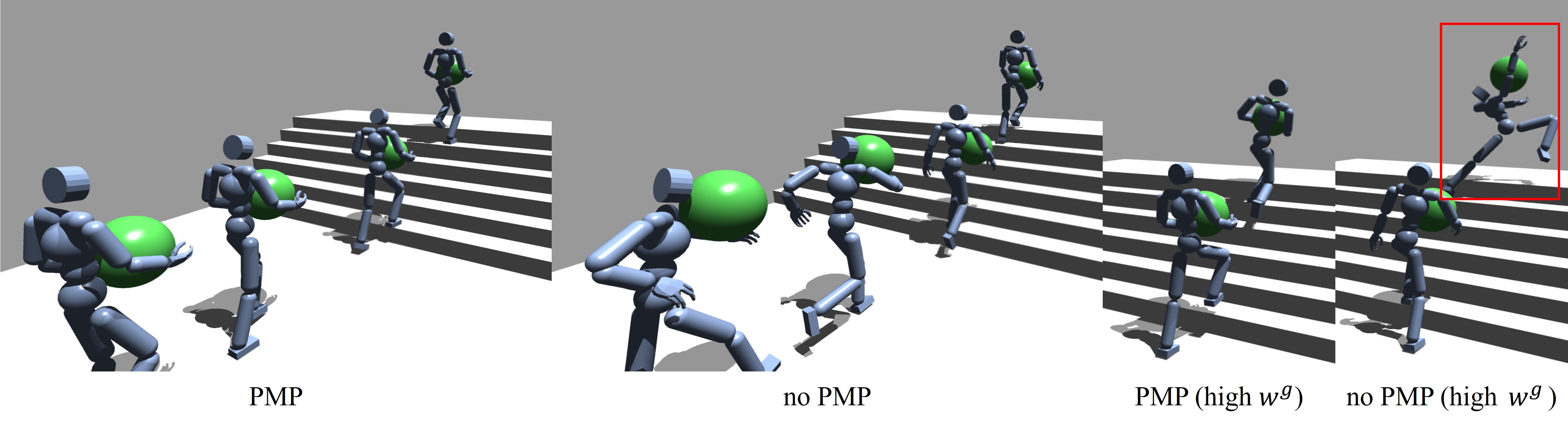}
  \caption{Comparison between PMP and no PMP in \textit{Upstair Carrying}. Note agent trained without PMP loses naturalness in the higher task reward setting.}
  \label{fig:appendix_staircarry}
\end{figure*}

\subsection{Results}\label{sec:results}
We first show the quantitative results of our experiments.
To measure the task performance, we evaluate a normalized average returns (NR) of the task reward as in \cite{2018-TOG-deepMimic}.
Similar to the previous work, we also enable the early termination and reference state initialization~\cite{2018-TOG-deepMimic} during the measurements, and expand the maximum length of an episode for some scenarios to show the stability in a long-term performance.
We average the values throughout 1000 episodes per experiment.
Although the maximum value of NR is inherently unreachable for some scenarios, extremely low NR ($<0.05$) implies complete failure in the training.
For \textit{Rope Climbing}, we additionally test agents climbing a rope with a lower frictional coefficient to show the effect of PMP.

\begin{figure*}[h]
  \includegraphics[width=\textwidth]{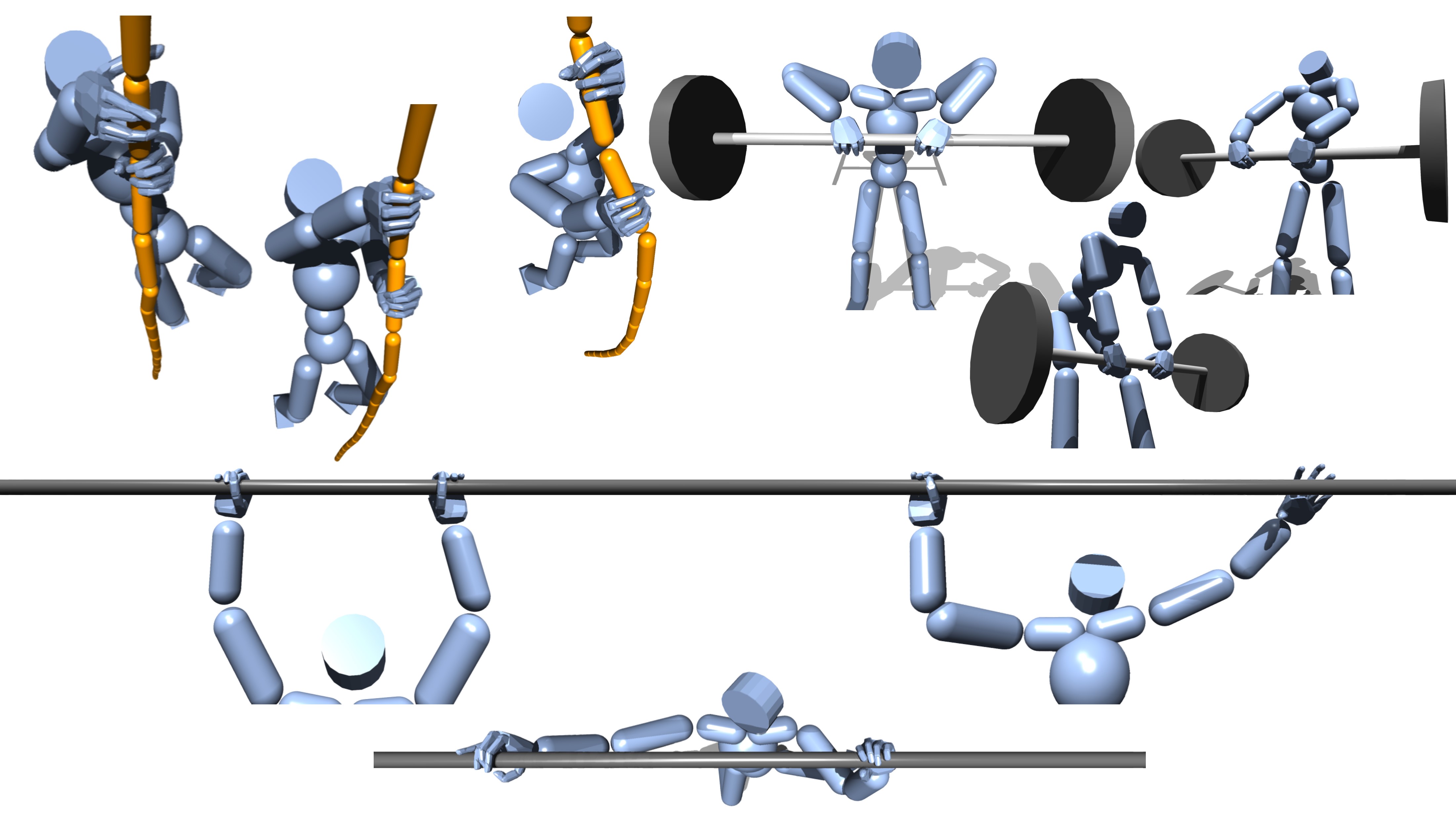}
  \caption{Snapshots on the grasping moments in \textit{Rope Climbing} (up left), \textit{Barbell Lifting} (up right), and \textit{Bar Hanging} (bottom).}
  \label{fig:appendix_grasp_motions}
\end{figure*}

As shown in Table~\ref{tab:comparison}, our approach outperforms the agents without PMP for all the scenarios.
This implies that in the perspective of task efficiency, our method can extract motion priors more effectively from the same amount of motion captures compared to the baseline.
Moreover, PMP not only enhances task performance but also generates natural motions that resembles human-object interaction in the real world.
We additionally exhibit qualitative results of agents performing in the scenarios as shown in Figure~\ref{fig:interaction_scenes} and ~\ref{fig:appendix_grasp_motions}.
We further provide results on the other examples including comparisons with the baselines in the supplementary \href{https://youtu.be/WdLGvKdNG-0}{\color{blue}video}.

\section{Discussion}\label{sec:discussion}

\begin{figure*}[h]
  \includegraphics[width=\textwidth]{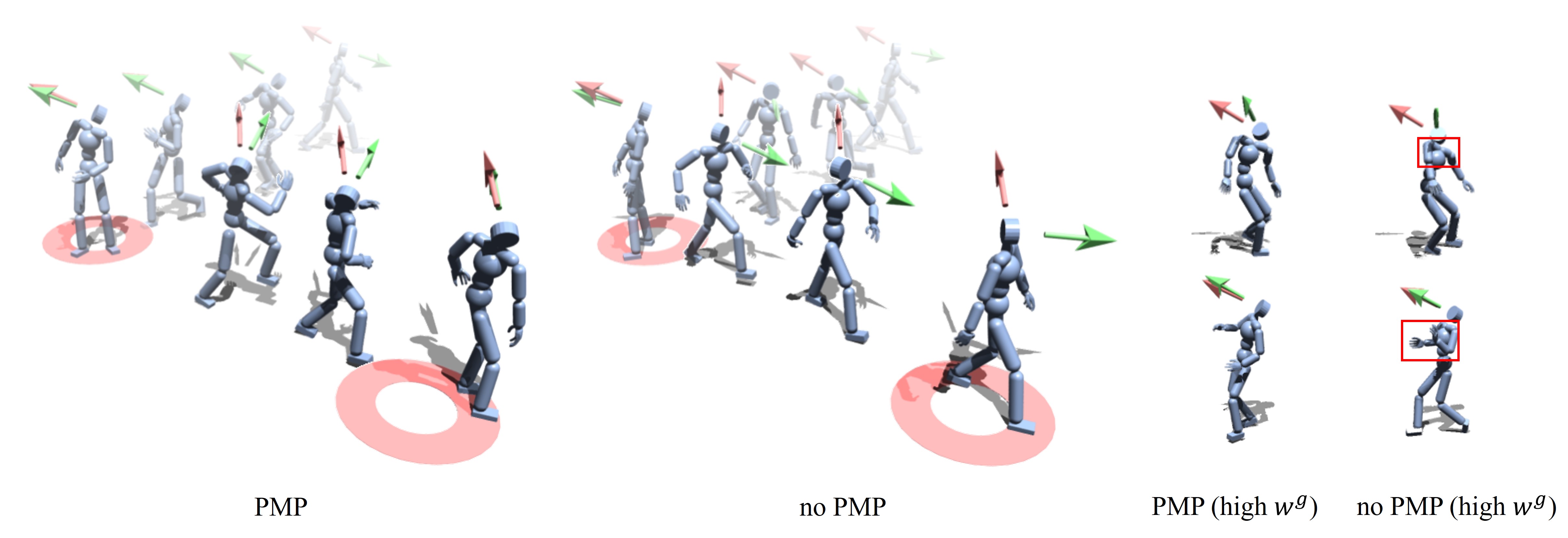}
  \caption{Comparison between PMP and no PMP in \textit{Sight Locomotion}. Red and green arrow indicate direction of goal and current sight respectively and the red marker in the ground represents target position. Similar to \textit{Upstair Carrying}, higher priority on the task reward decreases motion quality from agent trained without PMP.}
  \label{fig:appendix_locosight}
\end{figure*}

\paragraph{Skill Assembly}
We first describe the strength of our method when the number of available motion captures is limited using the results of \textit{Upstair Carrying} and \textit{Sight Locomotion}.
In Figure~\ref{fig:locosight_staircarry_graph}, we evaluate two task-related measurements: 
the joint position error in \textit{Upstair Carrying} compares the $L2$ distance of the upper body poses compared to the \textit{carrying idle} pose, and the sight tracking error measures the angle difference between the actual sight direction and the goal direction for \textit{Sight Locomotion}.
In both scenarios, our method synthesizes meaningful skills through the combinations of existing part-wise motion data.
However in \textit{Upstair Carrying}, the agent without PMP treats locomotion as a more important skill to reach the goal, and thus it fails to imitate carrying motion.
Similarly in \textit{Sight Locomotion}, our approach shows agile sight transitions whenever the sight goal changes while the baseline results in larger errors.
These results are the direct consequences of the design of the algorithm, where the policy without PMP refers to a particular full-body motion among the reference sequences, while our approach explores more rich set of skills to find the best combination for the task.
We find that even with a higher weight for task reward, the agent trained with PMP maintains natural motion whereas the non-PMP agent sacrifices the quality of motions.
Further visualizations of the qualitative results of the two scenarios are shown in Figure~\ref{fig:appendix_staircarry} and~\ref{fig:appendix_locosight}.

\begin{figure*}[h]
  \includegraphics[width=0.95\textwidth]{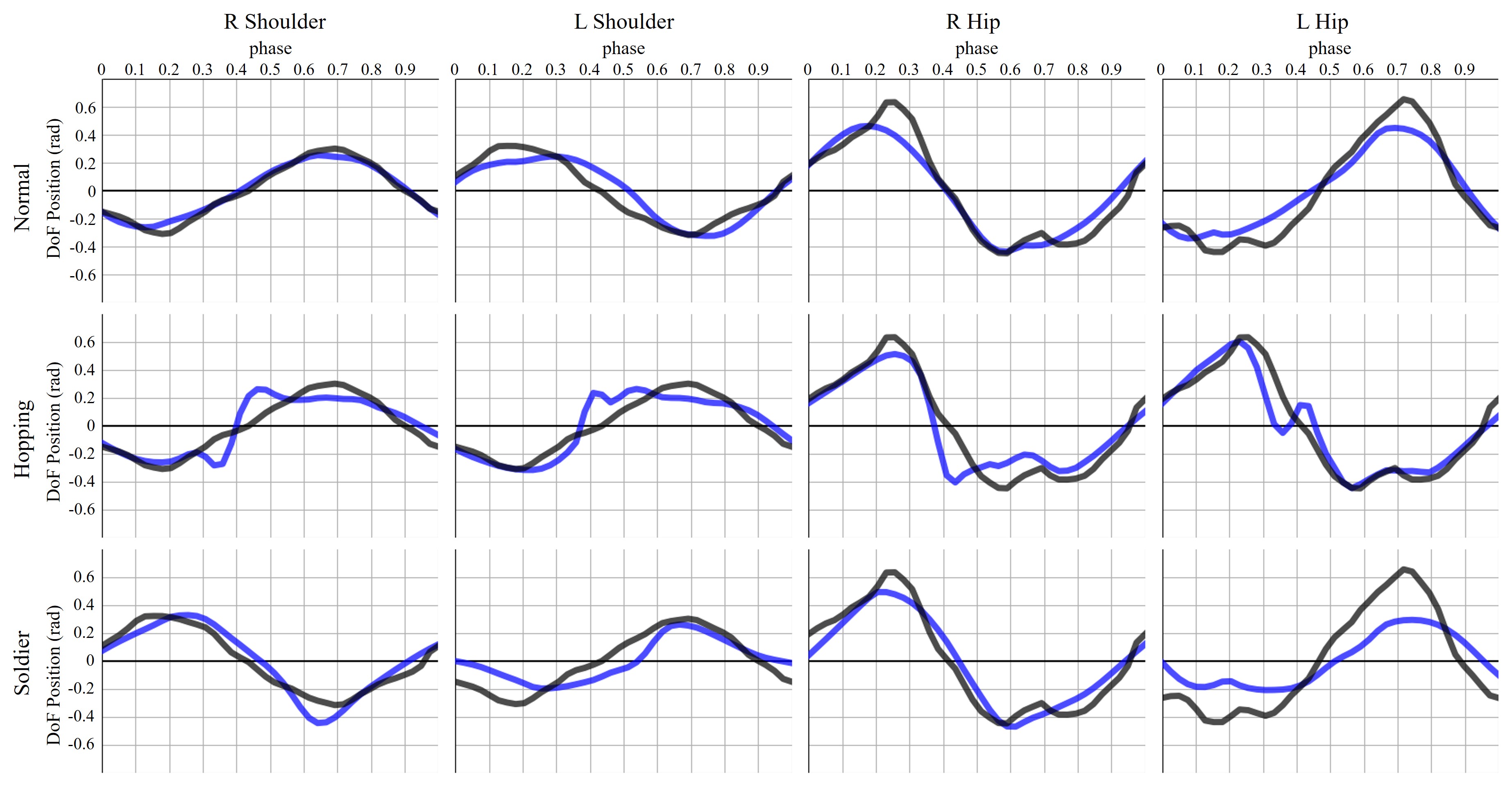}
  \caption{4 DoF positions (along the $y$ axis of right and left shoulders and hips) of PMP agent (blue) compared with the reference motions (black) for various walking styles. Note speed of PMP agent's walking is scaled to match phase of reference motions.}
  \label{fig:appendix_walking_phases}
\end{figure*}

\paragraph{Motion Augmentation}
In \textit{Walking Styles}, we experiment on the capacity of PMP on the motion style augmentation.
By controlling the temporal phases for the shoulders and hips only as an additional motion prior, our agent performs different styles of natural walking in a physically plausible manner.
The overall results on the tracking quality of each style are shown in Figure~\ref{fig:appendix_walking_phases}.
Among the experiments, we find the demo blend technique (Sec.~\ref{sec:demo_blend}) is critical in imitating the hopping-style motion.
Since we train 8 discriminators in total for this example, an agent is highly susceptible to the reward vanishing problem.
Figure~\ref{fig:demo_blend_result} qualitatively compares the effect of demo blend technique.
The demo blending maintains the style rewards in effect such that all the parts observable while training.

\paragraph {Generalization of Interaction Prior}
Even though the agent in the interaction gym (Sec.~\ref{sec:interaction_gym}) only experiences grasping of a simple cylindrical rod, target objects in the

\newpage
actual interactions can have more diverse shapes.
In the sub-scenarios of \textit{Cart Pulling}, we test whether interaction prior can be generalized to different types of handles or various grips.
In Figure~\ref{fig:pullcart_shapes}, we find that the interaction prior can be generalized to a thicker handle (3.5 cm radius) as well as a completely different shape like curved target.
Furthermore, we motivate the agent to use various grasping styles by simply changing the interaction target to different parts of the handle.
We find that the simple modification results in different grips, and consequently affects the whole-body motion.
This shows that users can diversify the details of the targeting scenario using an interaction prior obtained from a single training.

\paragraph{Whole-body Naturalness}
Our pipeline effectively balances part-wise motions without explicit module for global coordination.
We attribute this to the physics simulator and the reward term.
The simulator only allows physically valid whole-body motion and filters out combinations that result in awkward postures.
Additionally, the task reward encourages the agent to efficiently achieve the goal and suppresses irrelevant movements.
The reward in some examples also contain torque minimizing rewards, which contribute to the smoother overall trajectories.

\paragraph {Limitation}
PMP can provide a great option when motion captures are limited or interaction prior for a specific part is required.
Also, proposed interaction prior reduces repetitive policy optimization for the grasping-related reward terms in multiple downstream tasks.
However still, proposed method does not completely relieve all the efforts on inventing an effective task reward function to induce agent to properly behave in the challenging scenarios.
As a future work, PMP may include additional demo features, such as visual inputs of the interaction scenes which may enable the agent to automatically explore the optimal combination of the part-wise skills with comparably simpler task reward terms.
In addition, an automated process of exploring the optimal part segment combination can be addressed in the succeeding work.

\section{Conclusion}
In this work, we present a framework to incorporate part-wise motion priors to assemble multiple motion skills for a whole-body agent.
We segment the body joints into parts and encourage each part to refer to different demonstrations.
In our framework, an agent can benefit from both motion captures and simulated trajectories for a specific subpart of the body, which is demonstrated with the pre-trained grasping skill learned from a minimal hand-only gym.
Our approach composes a complete skill of physical interaction by a novel combination of motions that is not available in the motion capture, and enables agent to perform in the various challenging tasks.
A range of scenarios can be broadened with another choice of the dexterous part and interaction skill, \textit{e.g.} toes with the sophisticated control of a soccer ball.
We believe that the our method can be further expanded to animations with scene-level interaction or robotic whole-body manipulators in contact-rich environments.

%%
%% The acknowledgments section is defined using the "acks" environment
%% (and NOT an unnumbered section). This ensures the proper
%% identification of the section in the article metadata, and the
%% consistent spelling of the heading.
\begin{acks}
This work was supported by the National Research Foundation of Korea(NRF) grant funded by the Korea government(MSIT) (No. RS-2023-00208197) and Creative-Pioneering Researchers Program through Seoul National University. Jungdam Won was partially supported by the MSIT(Ministry of Science and ICT), Korea, under the ITRC(Information Technology Research Center) support program(IITP-2023-2020-0-01460) supervised by the IITP(Institute for Information \& Communications Technology Planning \& Evaluation), and by ICT (Institute of Computer Technology) at Seoul National University.
\end{acks}

%%
%% The next two lines define the bibliography style to be used, and
%% the bibliography file.
% \newpage
\bibliographystyle{ACM-Reference-Format}
\bibliography{sample-base}

%%% -*-BibTeX-*-
%%% Do NOT edit. File created by BibTeX with style
%%% ACM-Reference-Format-Journals [18-Jan-2012].

\begin{thebibliography}{42}

%%% ====================================================================
%%% NOTE TO THE USER: you can override these defaults by providing
%%% customized versions of any of these macros before the \bibliography
%%% command.  Each of them MUST provide its own final punctuation,
%%% except for \shownote{}, \showDOI{}, and \showURL{}.  The latter two
%%% do not use final punctuation, in order to avoid confusing it with
%%% the Web address.
%%%
%%% To suppress output of a particular field, define its macro to expand
%%% to an empty string, or better, \unskip, like this:
%%%
%%% \newcommand{\showDOI}[1]{\unskip}   % LaTeX syntax
%%%
%%% \def \showDOI #1{\unskip}           % plain TeX syntax
%%%
%%% ====================================================================

\ifx \showCODEN    \undefined \def \showCODEN     #1{\unskip}     \fi
\ifx \showDOI      \undefined \def \showDOI       #1{#1}\fi
\ifx \showISBNx    \undefined \def \showISBNx     #1{\unskip}     \fi
\ifx \showISBNxiii \undefined \def \showISBNxiii  #1{\unskip}     \fi
\ifx \showISSN     \undefined \def \showISSN      #1{\unskip}     \fi
\ifx \showLCCN     \undefined \def \showLCCN      #1{\unskip}     \fi
\ifx \shownote     \undefined \def \shownote      #1{#1}          \fi
\ifx \showarticletitle \undefined \def \showarticletitle #1{#1}   \fi
\ifx \showURL      \undefined \def \showURL       {\relax}        \fi
% The following commands are used for tagged output and should be
% invisible to TeX
\providecommand\bibfield[2]{#2}
\providecommand\bibinfo[2]{#2}
\providecommand\natexlab[1]{#1}
\providecommand\showeprint[2][]{arXiv:#2}

\bibitem[Adobe(2020)]%
        {mixamo}
\bibfield{author}{\bibinfo{person}{Adobe}.} \bibinfo{year}{2020}\natexlab{}.
\newblock \bibinfo{booktitle}{\emph{Adobe's Mixamo}}.
\newblock Adobe.
\newblock
\urldef\tempurl%
\url{http://www.mixamo.com}
\showURL{%
\tempurl}


\bibitem[Bansal et~al\mbox{.}(2017)]%
        {bansal2017emergent}
\bibfield{author}{\bibinfo{person}{Trapit Bansal}, \bibinfo{person}{Jakub
  Pachocki}, \bibinfo{person}{Szymon Sidor}, \bibinfo{person}{Ilya Sutskever},
  {and} \bibinfo{person}{Igor Mordatch}.} \bibinfo{year}{2017}\natexlab{}.
\newblock \showarticletitle{Emergent complexity via multi-agent competition}.
\newblock \bibinfo{journal}{\emph{arXiv preprint arXiv:1710.03748}}
  (\bibinfo{year}{2017}).
\newblock


\bibitem[Bergamin et~al\mbox{.}(2019)]%
        {bergamin2019drecon}
\bibfield{author}{\bibinfo{person}{Kevin Bergamin}, \bibinfo{person}{Simon
  Clavet}, \bibinfo{person}{Daniel Holden}, {and}
  \bibinfo{person}{James~Richard Forbes}.} \bibinfo{year}{2019}\natexlab{}.
\newblock \showarticletitle{DReCon: data-driven responsive control of
  physics-based characters}.
\newblock \bibinfo{journal}{\emph{ACM Transactions On Graphics (TOG)}}
  \bibinfo{volume}{38}, \bibinfo{number}{6} (\bibinfo{year}{2019}),
  \bibinfo{pages}{1--11}.
\newblock


\bibitem[Fussell et~al\mbox{.}(2021)]%
        {fussell2021supertrack}
\bibfield{author}{\bibinfo{person}{Levi Fussell}, \bibinfo{person}{Kevin
  Bergamin}, {and} \bibinfo{person}{Daniel Holden}.}
  \bibinfo{year}{2021}\natexlab{}.
\newblock \showarticletitle{Supertrack: Motion tracking for physically
  simulated characters using supervised learning}.
\newblock \bibinfo{journal}{\emph{ACM Transactions on Graphics (TOG)}}
  \bibinfo{volume}{40}, \bibinfo{number}{6} (\bibinfo{year}{2021}),
  \bibinfo{pages}{1--13}.
\newblock


\bibitem[Ghosh et~al\mbox{.}(2022)]%
        {ghosh2022imos}
\bibfield{author}{\bibinfo{person}{Anindita Ghosh}, \bibinfo{person}{Rishabh
  Dabral}, \bibinfo{person}{Vladislav Golyanik}, \bibinfo{person}{Christian
  Theobalt}, {and} \bibinfo{person}{Philipp Slusallek}.}
  \bibinfo{year}{2022}\natexlab{}.
\newblock \showarticletitle{IMoS: Intent-Driven Full-Body Motion Synthesis for
  Human-Object Interactions}.
\newblock \bibinfo{journal}{\emph{arXiv preprint arXiv:2212.07555}}
  (\bibinfo{year}{2022}).
\newblock


\bibitem[Harvey et~al\mbox{.}(2020)]%
        {harvey2020robust}
\bibfield{author}{\bibinfo{person}{F{\'e}lix~G Harvey}, \bibinfo{person}{Mike
  Yurick}, \bibinfo{person}{Derek Nowrouzezahrai}, {and}
  \bibinfo{person}{Christopher Pal}.} \bibinfo{year}{2020}\natexlab{}.
\newblock \showarticletitle{Robust motion in-betweening}.
\newblock \bibinfo{journal}{\emph{ACM Transactions on Graphics (TOG)}}
  \bibinfo{volume}{39}, \bibinfo{number}{4} (\bibinfo{year}{2020}),
  \bibinfo{pages}{60--1}.
\newblock


\bibitem[Hasenclever et~al\mbox{.}(2020)]%
        {hasenclever2020comic}
\bibfield{author}{\bibinfo{person}{Leonard Hasenclever}, \bibinfo{person}{Fabio
  Pardo}, \bibinfo{person}{Raia Hadsell}, \bibinfo{person}{Nicolas Heess},
  {and} \bibinfo{person}{Josh Merel}.} \bibinfo{year}{2020}\natexlab{}.
\newblock \showarticletitle{Comic: Complementary task learning \& mimicry for
  reusable skills}. In \bibinfo{booktitle}{\emph{International Conference on
  Machine Learning}}. PMLR, \bibinfo{pages}{4105--4115}.
\newblock


\bibitem[Hassan et~al\mbox{.}(2021)]%
        {hassan2021populating}
\bibfield{author}{\bibinfo{person}{Mohamed Hassan}, \bibinfo{person}{Partha
  Ghosh}, \bibinfo{person}{Joachim Tesch}, \bibinfo{person}{Dimitrios Tzionas},
  {and} \bibinfo{person}{Michael~J Black}.} \bibinfo{year}{2021}\natexlab{}.
\newblock \showarticletitle{Populating 3D scenes by learning human-scene
  interaction}. In \bibinfo{booktitle}{\emph{Proceedings of the IEEE/CVF
  Conference on Computer Vision and Pattern Recognition}}.
  \bibinfo{pages}{14708--14718}.
\newblock


\bibitem[Hecker et~al\mbox{.}(2008)]%
        {hecker2008real}
\bibfield{author}{\bibinfo{person}{Chris Hecker}, \bibinfo{person}{Bernd
  Raabe}, \bibinfo{person}{Ryan~W Enslow}, \bibinfo{person}{John DeWeese},
  \bibinfo{person}{Jordan Maynard}, {and} \bibinfo{person}{Kees van Prooijen}.}
  \bibinfo{year}{2008}\natexlab{}.
\newblock \showarticletitle{Real-time motion retargeting to highly varied
  user-created morphologies}.
\newblock \bibinfo{journal}{\emph{ACM Transactions on Graphics (TOG)}}
  \bibinfo{volume}{27}, \bibinfo{number}{3} (\bibinfo{year}{2008}),
  \bibinfo{pages}{1--11}.
\newblock


\bibitem[Ho and Ermon(2016)]%
        {ho2016generative}
\bibfield{author}{\bibinfo{person}{Jonathan Ho} {and} \bibinfo{person}{Stefano
  Ermon}.} \bibinfo{year}{2016}\natexlab{}.
\newblock \showarticletitle{Generative adversarial imitation learning}.
\newblock \bibinfo{journal}{\emph{Advances in neural information processing
  systems}}  \bibinfo{volume}{29} (\bibinfo{year}{2016}).
\newblock


\bibitem[Holden et~al\mbox{.}(2017)]%
        {holden2017phase}
\bibfield{author}{\bibinfo{person}{Daniel Holden}, \bibinfo{person}{Taku
  Komura}, {and} \bibinfo{person}{Jun Saito}.} \bibinfo{year}{2017}\natexlab{}.
\newblock \showarticletitle{Phase-functioned neural networks for character
  control}.
\newblock \bibinfo{journal}{\emph{ACM Transactions on Graphics (TOG)}}
  \bibinfo{volume}{36}, \bibinfo{number}{4} (\bibinfo{year}{2017}),
  \bibinfo{pages}{1--13}.
\newblock


\bibitem[Jang et~al\mbox{.}(2022)]%
        {jang2022motion}
\bibfield{author}{\bibinfo{person}{Deok-Kyeong Jang}, \bibinfo{person}{Soomin
  Park}, {and} \bibinfo{person}{Sung-Hee Lee}.}
  \bibinfo{year}{2022}\natexlab{}.
\newblock \showarticletitle{Motion Puzzle: Arbitrary Motion Style Transfer by
  Body Part}.
\newblock \bibinfo{journal}{\emph{ACM Transactions on Graphics (TOG)}}
  (\bibinfo{year}{2022}).
\newblock


\bibitem[Kumar and Todorov(2015)]%
        {kumar2015mujoco}
\bibfield{author}{\bibinfo{person}{Vikash Kumar} {and} \bibinfo{person}{Emanuel
  Todorov}.} \bibinfo{year}{2015}\natexlab{}.
\newblock \showarticletitle{Mujoco haptix: A virtual reality system for hand
  manipulation}. In \bibinfo{booktitle}{\emph{2015 IEEE-RAS 15th International
  Conference on Humanoid Robots (Humanoids)}}. IEEE, \bibinfo{pages}{657--663}.
\newblock


\bibitem[Lee et~al\mbox{.}(2006)]%
        {lee2006motion}
\bibfield{author}{\bibinfo{person}{Kang~Hoon Lee}, \bibinfo{person}{Myung~Geol
  Choi}, {and} \bibinfo{person}{Jehee Lee}.} \bibinfo{year}{2006}\natexlab{}.
\newblock \showarticletitle{Motion patches: building blocks for virtual
  environments annotated with motion data}.
\newblock In \bibinfo{booktitle}{\emph{ACM SIGGRAPH 2006 Papers}}.
  \bibinfo{pages}{898--906}.
\newblock


\bibitem[Lee et~al\mbox{.}(2022)]%
        {lee2022learning}
\bibfield{author}{\bibinfo{person}{Seyoung Lee}, \bibinfo{person}{Jiye Lee},
  {and} \bibinfo{person}{Jehee Lee}.} \bibinfo{year}{2022}\natexlab{}.
\newblock \showarticletitle{Learning Virtual Chimeras by Dynamic Motion
  Reassembly}.
\newblock \bibinfo{journal}{\emph{ACM Transactions on Graphics (TOG)}}
  \bibinfo{volume}{41}, \bibinfo{number}{6} (\bibinfo{year}{2022}),
  \bibinfo{pages}{1--13}.
\newblock


\bibitem[Liu and Hodgins(2018)]%
        {liu2018learning}
\bibfield{author}{\bibinfo{person}{Libin Liu} {and} \bibinfo{person}{Jessica
  Hodgins}.} \bibinfo{year}{2018}\natexlab{}.
\newblock \showarticletitle{Learning basketball dribbling skills using
  trajectory optimization and deep reinforcement learning}.
\newblock \bibinfo{journal}{\emph{ACM Transactions on Graphics (TOG)}}
  \bibinfo{volume}{37}, \bibinfo{number}{4} (\bibinfo{year}{2018}),
  \bibinfo{pages}{1--14}.
\newblock


\bibitem[Makoviychuk et~al\mbox{.}(2021)]%
        {makoviychuk2021isaac}
\bibfield{author}{\bibinfo{person}{Viktor Makoviychuk}, \bibinfo{person}{Lukasz
  Wawrzyniak}, \bibinfo{person}{Yunrong Guo}, \bibinfo{person}{Michelle Lu},
  \bibinfo{person}{Kier Storey}, \bibinfo{person}{Miles Macklin},
  \bibinfo{person}{David Hoeller}, \bibinfo{person}{Nikita Rudin},
  \bibinfo{person}{Arthur Allshire}, \bibinfo{person}{Ankur Handa}, {and}
  \bibinfo{person}{Gavriel State}.} \bibinfo{year}{2021}\natexlab{}.
\newblock \bibinfo{title}{Isaac Gym: High Performance GPU-Based Physics
  Simulation For Robot Learning}.
\newblock
\newblock


\bibitem[Merel et~al\mbox{.}(2020)]%
        {merel2020catch}
\bibfield{author}{\bibinfo{person}{Josh Merel}, \bibinfo{person}{Saran
  Tunyasuvunakool}, \bibinfo{person}{Arun Ahuja}, \bibinfo{person}{Yuval
  Tassa}, \bibinfo{person}{Leonard Hasenclever}, \bibinfo{person}{Vu Pham},
  \bibinfo{person}{Tom Erez}, \bibinfo{person}{Greg Wayne}, {and}
  \bibinfo{person}{Nicolas Heess}.} \bibinfo{year}{2020}\natexlab{}.
\newblock \showarticletitle{Catch \& Carry: reusable neural controllers for
  vision-guided whole-body tasks}.
\newblock \bibinfo{journal}{\emph{ACM Transactions on Graphics (TOG)}}
  \bibinfo{volume}{39}, \bibinfo{number}{4} (\bibinfo{year}{2020}),
  \bibinfo{pages}{39--1}.
\newblock


\bibitem[Park et~al\mbox{.}(2022)]%
        {park2022handoccnet}
\bibfield{author}{\bibinfo{person}{JoonKyu Park}, \bibinfo{person}{Yeonguk Oh},
  \bibinfo{person}{Gyeongsik Moon}, \bibinfo{person}{Hongsuk Choi}, {and}
  \bibinfo{person}{Kyoung~Mu Lee}.} \bibinfo{year}{2022}\natexlab{}.
\newblock \showarticletitle{HandOccNet: Occlusion-Robust 3D Hand Mesh
  Estimation Network}. In \bibinfo{booktitle}{\emph{Proceedings of the IEEE/CVF
  Conference on Computer Vision and Pattern Recognition}}.
  \bibinfo{pages}{1496--1505}.
\newblock


\bibitem[Park et~al\mbox{.}(2019)]%
        {park2019learning}
\bibfield{author}{\bibinfo{person}{Soohwan Park}, \bibinfo{person}{Hoseok Ryu},
  \bibinfo{person}{Seyoung Lee}, \bibinfo{person}{Sunmin Lee}, {and}
  \bibinfo{person}{Jehee Lee}.} \bibinfo{year}{2019}\natexlab{}.
\newblock \showarticletitle{Learning predict-and-simulate policies from
  unorganized human motion data}.
\newblock \bibinfo{journal}{\emph{ACM Transactions on Graphics (TOG)}}
  \bibinfo{volume}{38}, \bibinfo{number}{6} (\bibinfo{year}{2019}),
  \bibinfo{pages}{1--11}.
\newblock


\bibitem[Peng et~al\mbox{.}(2018)]%
        {2018-TOG-deepMimic}
\bibfield{author}{\bibinfo{person}{Xue~Bin Peng}, \bibinfo{person}{Pieter
  Abbeel}, \bibinfo{person}{Sergey Levine}, {and} \bibinfo{person}{Michiel
  van~de Panne}.} \bibinfo{year}{2018}\natexlab{}.
\newblock \showarticletitle{DeepMimic: Example-guided Deep Reinforcement
  Learning of Physics-based Character Skills}.
\newblock \bibinfo{journal}{\emph{ACM Trans. Graph.}} \bibinfo{volume}{37},
  \bibinfo{number}{4}, Article \bibinfo{articleno}{143} (\bibinfo{date}{July}
  \bibinfo{year}{2018}), \bibinfo{numpages}{14}~pages.
\newblock
\showISSN{0730-0301}
\urldef\tempurl%
\url{https://doi.org/10.1145/3197517.3201311}
\showDOI{\tempurl}


\bibitem[Peng et~al\mbox{.}(2022)]%
        {peng2022ase}
\bibfield{author}{\bibinfo{person}{Xue~Bin Peng}, \bibinfo{person}{Yunrong
  Guo}, \bibinfo{person}{Lina Halper}, \bibinfo{person}{Sergey Levine}, {and}
  \bibinfo{person}{Sanja Fidler}.} \bibinfo{year}{2022}\natexlab{}.
\newblock \showarticletitle{Ase: Large-scale reusable adversarial skill
  embeddings for physically simulated characters}.
\newblock \bibinfo{journal}{\emph{ACM Transactions On Graphics (TOG)}}
  \bibinfo{volume}{41}, \bibinfo{number}{4} (\bibinfo{year}{2022}),
  \bibinfo{pages}{1--17}.
\newblock


\bibitem[Peng et~al\mbox{.}(2021)]%
        {peng2021amp}
\bibfield{author}{\bibinfo{person}{Xue~Bin Peng}, \bibinfo{person}{Ze Ma},
  \bibinfo{person}{Pieter Abbeel}, \bibinfo{person}{Sergey Levine}, {and}
  \bibinfo{person}{Angjoo Kanazawa}.} \bibinfo{year}{2021}\natexlab{}.
\newblock \showarticletitle{Amp: Adversarial motion priors for stylized
  physics-based character control}.
\newblock \bibinfo{journal}{\emph{ACM Transactions on Graphics (TOG)}}
  \bibinfo{volume}{40}, \bibinfo{number}{4} (\bibinfo{year}{2021}),
  \bibinfo{pages}{1--20}.
\newblock


\bibitem[Petrovich et~al\mbox{.}(2021)]%
        {petrovich2021action}
\bibfield{author}{\bibinfo{person}{Mathis Petrovich},
  \bibinfo{person}{Michael~J Black}, {and} \bibinfo{person}{G{\"u}l Varol}.}
  \bibinfo{year}{2021}\natexlab{}.
\newblock \showarticletitle{Action-conditioned 3d human motion synthesis with
  transformer vae}. In \bibinfo{booktitle}{\emph{Proceedings of the IEEE/CVF
  International Conference on Computer Vision}}. \bibinfo{pages}{10985--10995}.
\newblock


\bibitem[Schulman et~al\mbox{.}(2017)]%
        {schulman2017proximal}
\bibfield{author}{\bibinfo{person}{John Schulman}, \bibinfo{person}{Filip
  Wolski}, \bibinfo{person}{Prafulla Dhariwal}, \bibinfo{person}{Alec Radford},
  {and} \bibinfo{person}{Oleg Klimov}.} \bibinfo{year}{2017}\natexlab{}.
\newblock \showarticletitle{Proximal policy optimization algorithms}.
\newblock \bibinfo{journal}{\emph{arXiv preprint arXiv:1707.06347}}
  (\bibinfo{year}{2017}).
\newblock


\bibitem[Shum et~al\mbox{.}(2008)]%
        {shum2008simulating}
\bibfield{author}{\bibinfo{person}{Hubert~PH Shum}, \bibinfo{person}{Taku
  Komura}, {and} \bibinfo{person}{Shuntaro Yamazaki}.}
  \bibinfo{year}{2008}\natexlab{}.
\newblock \showarticletitle{Simulating interactions of avatars in high
  dimensional state space}. In \bibinfo{booktitle}{\emph{Proceedings of the
  2008 Symposium on interactive 3D Graphics and Games}}.
  \bibinfo{pages}{131--138}.
\newblock


\bibitem[Shum et~al\mbox{.}(2010)]%
        {shum2010simulating}
\bibfield{author}{\bibinfo{person}{Hubert~PH Shum}, \bibinfo{person}{Taku
  Komura}, {and} \bibinfo{person}{Shuntaro Yamazaki}.}
  \bibinfo{year}{2010}\natexlab{}.
\newblock \showarticletitle{Simulating multiple character interactions with
  collaborative and adversarial goals}.
\newblock \bibinfo{journal}{\emph{IEEE Transactions on Visualization and
  Computer Graphics}} \bibinfo{volume}{18}, \bibinfo{number}{5}
  (\bibinfo{year}{2010}), \bibinfo{pages}{741--752}.
\newblock


\bibitem[Starke et~al\mbox{.}(2019)]%
        {starke2019neural}
\bibfield{author}{\bibinfo{person}{Sebastian Starke}, \bibinfo{person}{He
  Zhang}, \bibinfo{person}{Taku Komura}, {and} \bibinfo{person}{Jun Saito}.}
  \bibinfo{year}{2019}\natexlab{}.
\newblock \showarticletitle{Neural state machine for character-scene
  interactions.}
\newblock \bibinfo{journal}{\emph{ACM Trans. Graph.}} \bibinfo{volume}{38},
  \bibinfo{number}{6} (\bibinfo{year}{2019}), \bibinfo{pages}{209--1}.
\newblock


\bibitem[Taheri et~al\mbox{.}(2022)]%
        {taheri2022goal}
\bibfield{author}{\bibinfo{person}{Omid Taheri}, \bibinfo{person}{Vasileios
  Choutas}, \bibinfo{person}{Michael~J Black}, {and} \bibinfo{person}{Dimitrios
  Tzionas}.} \bibinfo{year}{2022}\natexlab{}.
\newblock \showarticletitle{Goal: Generating 4d whole-body motion for
  hand-object grasping}. In \bibinfo{booktitle}{\emph{Proceedings of the
  IEEE/CVF Conference on Computer Vision and Pattern Recognition}}.
  \bibinfo{pages}{13263--13273}.
\newblock


\bibitem[Tang et~al\mbox{.}(2022)]%
        {tang2022real}
\bibfield{author}{\bibinfo{person}{Xiangjun Tang}, \bibinfo{person}{He Wang},
  \bibinfo{person}{Bo Hu}, \bibinfo{person}{Xu Gong}, \bibinfo{person}{Ruifan
  Yi}, \bibinfo{person}{Qilong Kou}, {and} \bibinfo{person}{Xiaogang Jin}.}
  \bibinfo{year}{2022}\natexlab{}.
\newblock \showarticletitle{Real-time Controllable Motion Transition for
  Characters}.
\newblock \bibinfo{journal}{\emph{arXiv preprint arXiv:2205.02540}}
  (\bibinfo{year}{2022}).
\newblock


\bibitem[Tendulkar et~al\mbox{.}(2022)]%
        {tendulkar2022flex}
\bibfield{author}{\bibinfo{person}{Purva Tendulkar},
  \bibinfo{person}{D{\'\i}dac Sur{\'\i}s}, {and} \bibinfo{person}{Carl
  Vondrick}.} \bibinfo{year}{2022}\natexlab{}.
\newblock \showarticletitle{FLEX: Full-Body Grasping Without Full-Body Grasps}.
\newblock \bibinfo{journal}{\emph{arXiv preprint arXiv:2211.11903}}
  (\bibinfo{year}{2022}).
\newblock


\bibitem[Wang et~al\mbox{.}(2022)]%
        {wang2022towards}
\bibfield{author}{\bibinfo{person}{Jingbo Wang}, \bibinfo{person}{Yu Rong},
  \bibinfo{person}{Jingyuan Liu}, \bibinfo{person}{Sijie Yan},
  \bibinfo{person}{Dahua Lin}, {and} \bibinfo{person}{Bo Dai}.}
  \bibinfo{year}{2022}\natexlab{}.
\newblock \showarticletitle{Towards Diverse and Natural Scene-aware 3D Human
  Motion Synthesis}. In \bibinfo{booktitle}{\emph{Proceedings of the IEEE/CVF
  Conference on Computer Vision and Pattern Recognition}}.
  \bibinfo{pages}{20460--20469}.
\newblock


\bibitem[Wang et~al\mbox{.}(2021)]%
        {wang2021synthesizing}
\bibfield{author}{\bibinfo{person}{Jiashun Wang}, \bibinfo{person}{Huazhe Xu},
  \bibinfo{person}{Jingwei Xu}, \bibinfo{person}{Sifei Liu}, {and}
  \bibinfo{person}{Xiaolong Wang}.} \bibinfo{year}{2021}\natexlab{}.
\newblock \showarticletitle{Synthesizing long-term 3d human motion and
  interaction in 3d scenes}. In \bibinfo{booktitle}{\emph{Proceedings of the
  IEEE/CVF Conference on Computer Vision and Pattern Recognition}}.
  \bibinfo{pages}{9401--9411}.
\newblock


\bibitem[Won et~al\mbox{.}(2020)]%
        {won2020scalable}
\bibfield{author}{\bibinfo{person}{Jungdam Won}, \bibinfo{person}{Deepak
  Gopinath}, {and} \bibinfo{person}{Jessica Hodgins}.}
  \bibinfo{year}{2020}\natexlab{}.
\newblock \showarticletitle{A scalable approach to control diverse behaviors
  for physically simulated characters}.
\newblock \bibinfo{journal}{\emph{ACM Transactions on Graphics (TOG)}}
  \bibinfo{volume}{39}, \bibinfo{number}{4} (\bibinfo{year}{2020}),
  \bibinfo{pages}{33--1}.
\newblock


\bibitem[Won et~al\mbox{.}(2021)]%
        {won2021control}
\bibfield{author}{\bibinfo{person}{Jungdam Won}, \bibinfo{person}{Deepak
  Gopinath}, {and} \bibinfo{person}{Jessica Hodgins}.}
  \bibinfo{year}{2021}\natexlab{}.
\newblock \showarticletitle{Control strategies for physically simulated
  characters performing two-player competitive sports}.
\newblock \bibinfo{journal}{\emph{ACM Transactions on Graphics (TOG)}}
  \bibinfo{volume}{40}, \bibinfo{number}{4} (\bibinfo{year}{2021}),
  \bibinfo{pages}{1--11}.
\newblock


\bibitem[Won et~al\mbox{.}(2022)]%
        {won2022physics}
\bibfield{author}{\bibinfo{person}{Jungdam Won}, \bibinfo{person}{Deepak
  Gopinath}, {and} \bibinfo{person}{Jessica Hodgins}.}
  \bibinfo{year}{2022}\natexlab{}.
\newblock \showarticletitle{Physics-based character controllers using
  conditional vaes}.
\newblock \bibinfo{journal}{\emph{ACM Transactions on Graphics (TOG)}}
  \bibinfo{volume}{41}, \bibinfo{number}{4} (\bibinfo{year}{2022}),
  \bibinfo{pages}{1--12}.
\newblock


\bibitem[Won et~al\mbox{.}(2014)]%
        {won2014generating}
\bibfield{author}{\bibinfo{person}{Jungdam Won}, \bibinfo{person}{Kyungho Lee},
  \bibinfo{person}{Carol O'Sullivan}, \bibinfo{person}{Jessica~K Hodgins},
  {and} \bibinfo{person}{Jehee Lee}.} \bibinfo{year}{2014}\natexlab{}.
\newblock \showarticletitle{Generating and ranking diverse multi-character
  interactions}.
\newblock \bibinfo{journal}{\emph{ACM Transactions on Graphics (TOG)}}
  \bibinfo{volume}{33}, \bibinfo{number}{6} (\bibinfo{year}{2014}),
  \bibinfo{pages}{1--12}.
\newblock


\bibitem[Wu et~al\mbox{.}(2022)]%
        {wu2022saga}
\bibfield{author}{\bibinfo{person}{Yan Wu}, \bibinfo{person}{Jiahao Wang},
  \bibinfo{person}{Yan Zhang}, \bibinfo{person}{Siwei Zhang},
  \bibinfo{person}{Otmar Hilliges}, \bibinfo{person}{Fisher Yu}, {and}
  \bibinfo{person}{Siyu Tang}.} \bibinfo{year}{2022}\natexlab{}.
\newblock \showarticletitle{Saga: Stochastic whole-body grasping with contact}.
  In \bibinfo{booktitle}{\emph{Computer Vision--ECCV 2022: 17th European
  Conference, Tel Aviv, Israel, October 23--27, 2022, Proceedings, Part VI}}.
  Springer, \bibinfo{pages}{257--274}.
\newblock


\bibitem[Yang et~al\mbox{.}(2022)]%
        {yang2022learning}
\bibfield{author}{\bibinfo{person}{Zeshi Yang}, \bibinfo{person}{Kangkang Yin},
  {and} \bibinfo{person}{Libin Liu}.} \bibinfo{year}{2022}\natexlab{}.
\newblock \showarticletitle{Learning to use chopsticks in diverse gripping
  styles}.
\newblock \bibinfo{journal}{\emph{ACM Transactions on Graphics (TOG)}}
  \bibinfo{volume}{41}, \bibinfo{number}{4} (\bibinfo{year}{2022}),
  \bibinfo{pages}{1--17}.
\newblock


\bibitem[Zhang et~al\mbox{.}(2018)]%
        {zhang2018mode}
\bibfield{author}{\bibinfo{person}{He Zhang}, \bibinfo{person}{Sebastian
  Starke}, \bibinfo{person}{Taku Komura}, {and} \bibinfo{person}{Jun Saito}.}
  \bibinfo{year}{2018}\natexlab{}.
\newblock \showarticletitle{Mode-adaptive neural networks for quadruped motion
  control}.
\newblock \bibinfo{journal}{\emph{ACM Transactions on Graphics (TOG)}}
  \bibinfo{volume}{37}, \bibinfo{number}{4} (\bibinfo{year}{2018}),
  \bibinfo{pages}{1--11}.
\newblock


\bibitem[Zhang et~al\mbox{.}(2021)]%
        {zhang2021manipnet}
\bibfield{author}{\bibinfo{person}{He Zhang}, \bibinfo{person}{Yuting Ye},
  \bibinfo{person}{Takaaki Shiratori}, {and} \bibinfo{person}{Taku Komura}.}
  \bibinfo{year}{2021}\natexlab{}.
\newblock \showarticletitle{ManipNet: Neural manipulation synthesis with a
  hand-object spatial representation}.
\newblock \bibinfo{journal}{\emph{ACM Transactions on Graphics (ToG)}}
  \bibinfo{volume}{40}, \bibinfo{number}{4} (\bibinfo{year}{2021}),
  \bibinfo{pages}{1--14}.
\newblock


\bibitem[Zhao et~al\mbox{.}(2020)]%
        {zhao2020bayesian}
\bibfield{author}{\bibinfo{person}{Rui Zhao}, \bibinfo{person}{Hui Su}, {and}
  \bibinfo{person}{Qiang Ji}.} \bibinfo{year}{2020}\natexlab{}.
\newblock \showarticletitle{Bayesian adversarial human motion synthesis}. In
  \bibinfo{booktitle}{\emph{Proceedings of the IEEE/CVF Conference on Computer
  Vision and Pattern Recognition}}. \bibinfo{pages}{6225--6234}.
\newblock


\end{thebibliography}

%%
%% If your work has an appendix, this is the place to put it.
\newpage
\appendix
\section{Adversarial Training}
In this section, we further explain the details of the adversarial training for the part-wise motion priors (PMP).
Though the style discriminators and the interaction discriminators are mainly trained to minimize the discriminator loss $\mathcal{L}_{disc}$, a few regularization terms are also involved to mitigate the unstable nature of the adversarial training.
For simplification, we denote $\{D_{\phi_k}\}_K$ to refer to all the discriminators in our pipeline, which allows us to rewrite the discriminator loss as 
\begin{equation}
    \label{eq:pred_loss}
  \mathcal{L}_{disc}=~{\frac{1}{K}}\sum_{k=1}^K\left\{\mathbb{E}_{\mathcal{M}_k}[\log{(D_{\phi_k})}] + \mathbb{E}_{\pi_\theta}[\log{(1 - D_{\phi_k}})]\right\}.
\end{equation}

One of the additional regularization terms is the gradient penalty introduced in the previous work~\cite{peng2021amp}, which is to prevent the discriminator to be deviated from the distribution of the original demonstrations.
We modify the original formula to incorporate a set of discriminators, which results in
\begin{equation}
    \label{eq:modified_gp}
    \mathcal{L}_{gp}={\frac{1}{K}}\sum_{k=1}^K\mathbb{E}_{\mathcal{M}_k}[{\|\nabla_{(o_k,{o_k}')} D_{\phi_k}(o_k,{o_k}')\|^2}].
\end{equation}

In the other regularization term, we use weight decay to avoid the overfitting of the discriminators to the limited reference data with the squared norm regularization $\mathcal{L}_{reg}$ as
\begin{equation}
    \label{eq:modified_reg}
    \mathcal{L}_{reg}={\frac{1}{K}}\sum_{k=1}^K\|\phi_k\|^2.
\end{equation}

In summary, the total loss function to train the discriminators can be summarized as a weighted sum of the aforementioned terms
\begin{equation}
    \label{eq:total_loss}
    \mathcal{L}=w_{disc}\cdot \mathcal{L}_{disc} + w_{gp}\cdot \mathcal{L}_{gp} + w_{reg}\cdot \mathcal{L}_{reg},
\end{equation}
where $w_{disc}=5,~w_{gp}=5,~w_{reg}=0.0001$.
\section{Training Details}\label{sec:training_details}
In this section, we describe the detailed settings of the training including states and rewards used for each conducted task.

\subsection{General Settings}
As addressed in the main paper, our character is a humanoid equipped with two MPL~\cite{kumar2015mujoco} hands.
To train a whole-body agent for all the tasks, we include the proprioceptive information as well as the root states in the state vector.
The basic state configuration is largely adopted from the previous work~\cite{2018-TOG-deepMimic}, where state vector $s$ can be represented as
\begin{equation}
    s = [h^r,~\dot{p}^r,~q^r,~\dot{q}^r,~q,~\dot{q},~p^e].
\end{equation}
Here, each of $h^r,~\dot{p}^r,~q^r,~\dot{q}^r$ represents the height, velocity, orientation, and angular velocity of the root, while $q,~\dot{q}$ stand for the position and the velocity of the character joint respectively.
Further, $p^e$ is the relative positions of the palms and feet with respect to the root in Cartesian space.
Note that the orientation of the root and the spherical body joints are encoded as a 6D representation of rotation composed of the tangent and normal vectors.
In the scenarios highlighting interactions (\textit{Cart Pulling}, \textit{Bar Hanging}, \textit{Barbell Lifting}, \textit{Rope Climbing}), we augment the state vector with the interaction state $s^i$ of each hand that describes the state of the hand-only agent introduced in the Interaction Gym (Sec. 4.1).

We append additional task-specific information to each task.
For the scenario of \textit{Upstair Carrying, Sight Locomotion, Cart Pulling}, where the task contains the mission of target location~\cite{peng2021amp}, we add an extra feature of the relative position of root respect to the goal position.
Similarly, the relative Cartesian position of the barbell from its goal is included in the state vector for the \textit{Barbell Lifting} scenario.
However in \textit{Bar Hanging} and \textit{Rope Climbing}, an additional feature to indicate input hand goal is unnecessary as changes in the grasping target positions are already observable by an interaction state.

For the interaction scenario where two hands are guided to simultaneously grasp the object, we slightly modify the original style reward formulation of Eq. (9) into 
\begin{equation}
    r^s = c\cdot\{1-\max(\sigma_r, \sigma_l)\}\cdot\prod_{n\in H}\sigma_n r^i_n\cdot\prod_{k\notin{H}} r^s_k.
%  \end{gathered}
\end{equation}\label{eq:ours_style_reward_modified}
In this way, both hands are trained to utilize interaction prior in a synchronized manner, making interaction much more visually natural.
Empirically, we find that adding a small amount of offset, \textit{e.g.} 0.3, to the interaction reward $r^i_n$ can prevent the policy from converging to utilize only the kinematic priors.

As explained in the main paper, we use Euclidean distance-based Gaussian kernel $\Phi(u)$ to measure the interaction coefficient $\sigma$ given the observation of interaction state $u$.
We assume a hand is ready to interact with the target if the distance between the wrist and the object is lower than $20~cm$.
By reflecting the assumption, the formula of $\Phi$ is written as follows:
\begin{equation}
    \Phi(u) =  
    \begin{cases}
      1 & \|p(u)\| \le 10~cm \\
      \exp[-\gamma\cdot\|p(u)\|^3] & \|p(u)\| > 10~cm
    \end{cases}
\end{equation}\label{eq:gaussian_kernel}
where $\|p(u)\|$ is the distance between the wrist and the object from the observation $u$, and the constant $\gamma=4000$.

\subsection{Task-related Settings}
\paragraph{Interaction Gym}
\begin{figure}[t]
  \centering
  \includegraphics[width=\linewidth]{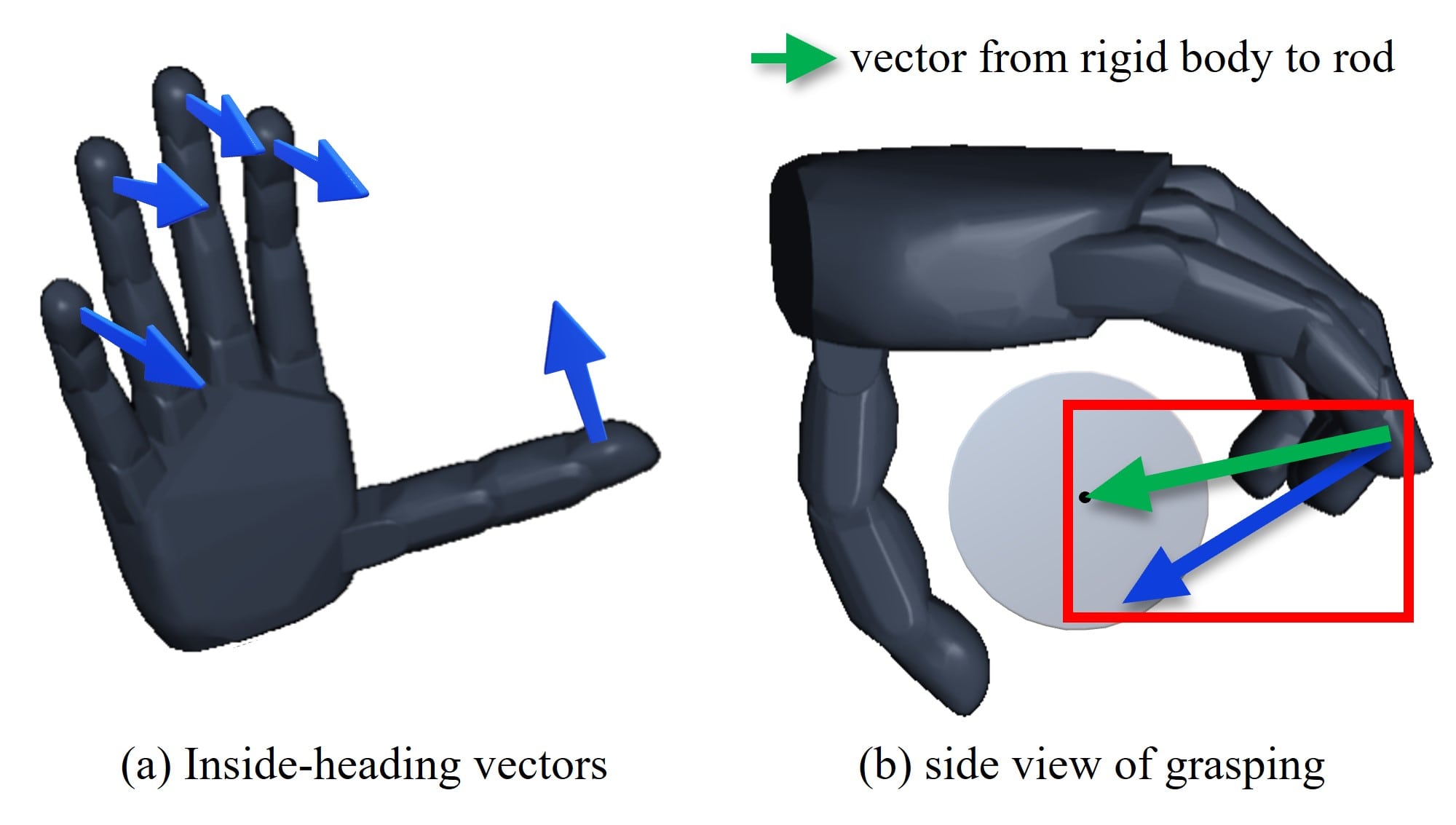}
  \caption{Visualization of the defined directional vectors in Interaction Gym. (a) heading direction $d_h$ toward the inside of the hand (blue), and (b) directional vector $d_r$ connecting each fingertip to the rod (green).}
  \label{fig:supplementary_vector_description}
\end{figure} 

In this environment, only the orientation of the rod is sampled randomly while its position is centered around the palm. This setting makes grasping always feasible without the translation of the root of the hand.
The force of $50\sim100~N$ and torque of $-30\sim30~N\cdot m$ in random directions are applied to the gravity-disabled $4~kg$ rod with the frequency of $30~Hz$.

To learn a stable yet natural grasping policy of a cylindrical rod against external disturbances, we introduce a set of reward terms in the interaction gym.
Along with the rod reward $r_{rod}$ motivating the firm grasping of the rod, we additionally devise several reward functions to reflect the grasping style of a real human hand.
First, finger reward $r_{fin}$ is designed to keep the rigid bodies comprising the hand close to the surface of the rod, and the MCP reward $r_{mcp}$ is applied to enforce MCP joints to exert maximal torques during grasping.
Here, a hand is guided from the tip reward $r_{tip}$ which minimizes the angle between heading direction $d_h$ toward the inside of the hand and the directional vector $d_r$ connecting each fingertip to the closest point on the surface of the target rod.
We visualize the $d_h$ and $d_r$ in Figure~\ref{fig:supplementary_vector_description}.
Next, to facilitate grasping in the various wrist pose, wrist reward $r_{wrist}$ is designed to track randomly set target pose while penalizing unnecessary wrist movements.
Lastly, actuated torque minimizing reward $r_{\tau}$ for DoFs except MCP joints is used to regularize hand motions.
The exact formulas for the aforementioned reward terms are written as follows:
\begin{equation}
    \label{eq:rod_reward}
    r_{rod} = 0.3\cdot \exp[-v_{rod}^2] + 0.7\cdot \exp[-0.1\cdot \omega_{rod}^2],
\end{equation}
\begin{equation}
    \label{eq:finger_reward}
    r_{fin} = \exp[-128\cdot \max_{k\in {K}} \|p_k\|^2],
\end{equation}
\begin{equation}
    \label{eq:mcp_reward}
    r_{mcp} = \exp[-3\cdot \|1-\bar{a}_{mcp}\|^2],
\end{equation}
\begin{equation}
    \label{eq:tip_reward}
    r_{tip} = \exp[-3\cdot \max_{k\in {K_{tip}}} \|1 - \langle d_{h,k}, d_{r,k}\rangle\|^2],
\end{equation}
\begin{equation}
    \label{eq:wrist_reward}
    r_{wrist} = \exp[-3\cdot\|\hat{q}_{wrist} - q_{wrist}\|^2]\cdot \exp[-0.1\cdot\|\dot{q}_{wrist}\|^2],
\end{equation}
\begin{equation}
    \label{eq:torque_reward}
    r_{\tau} = \exp[-0.002\cdot\sum_{j\notin J_{mcp}}\tau_j^2].
\end{equation}
In the equations, $v_{rod}$ and $\omega_{rod}$ refer to the linear and angular velocity of the rod, respectively. Further, $\bar{a}_{mcp}$ indicates a mean value of the normalized target action for the MCP joints which shapes hand at hard fist when $\bar{a}_{mcp}=1$.
In addition, each of $q_{wrist}, \dot{q}_{wrist}$ represents DoF position and velocity for the spherical wrist joint, and $\tau_j$ means actuated PD force of j-th joint.
Note that $K,~K_{tip}$ are a set of total hand, and fingertip rigid bodies respectively, and $J_{mcp}$ refer to the MCP joints set.
Consequently, the total task reward is calculated as
\begin{equation}
    \label{eq:total_hand_reward}
    r^g = 0.95\cdot r_{rod} \cdot r_{fin}\cdot r_{mcp}\cdot r_{tip}\cdot r_{wrist} + 0.05\cdot r_\tau.
\end{equation}

\paragraph{Upstair Carrying}
In this example, we build six stairs where each stair has a height of $0.2~m$ and width of $6~m$.
The task goal is to reach the goal position located at the highest stair, where the task reward is designed in a similar way to the target location~\cite{peng2021amp} task
\begin{equation}
    \label{eq:upstair_pos_reward}
    r_{pos} = \exp[-\gamma_{pos}\cdot\|p\|^2],
\end{equation}
\begin{equation}
    \label{eq:upstair_vel_reward}
    r_{vel} = \exp[-\gamma_{vel} \cdot \max(0, v^* - \langle v^r, {p\over\|p\|}\rangle)^2],
\end{equation}
\begin{equation}
    \label{eq:upstair_total_reward}
    r^g = w_{pos}\cdot r_{pos} + w_{vel}\cdot r_{vel}.
\end{equation}
Here, $p$ denotes the displacement from the root to the goal, and $v^r$ indicates the root velocity projected onto the heading direction.
Additionally, we set the target minimum speed $v^*=2~m/s$, and use $\gamma_{pos} = 0.5,~\gamma_{vel} = 1,~ w_{pos} = 0.7$,~$w_{vel}= 0.3$.

\paragraph{Sight Locomotion}
Similar to the \textit{Upstair Carrying} scenario, the task goal in this example also contains a target location where the goal is respawned $4~m$ far from the character root in the initial state.
Noteworthily, in this scenario, we have an additional goal for sight tracking, which is located on the surface of a cylinder with $1.5~m$ radius centered around the character.
We reuse the reward functions in Eq.~(\ref{eq:upstair_pos_reward}, \ref{eq:upstair_vel_reward}, \ref{eq:upstair_total_reward}) for root tracking reward $r_{root}$ with the same coefficients in \textit{Upstair Carrying}.
Additionally, we design a sight tracking reward, which is written as
\begin{equation}
    \label{eq:sigh_track_reward}
        r_{sight} = \exp[-\gamma_{sight}\cdot \|q_{goal} \ominus q_{head}\|^2],
\end{equation}
where each of $q_{goal}$ and $q_{head}$ represents the orientation of goal sight and current sight respectively, and the coefficient $\gamma_{sight}$ is $2$.
The total task reward is a weighted sum of $r_{root}$ and $r_{sight}$ with a ratio of 7:3.
\paragraph{Walking Styles}

This scenario is designed to show the style augmentation of a simple walking motion, where the task reward is adapted from the target heading~\cite{peng2021amp} task.
The reward function can be formulated similarly to the Eq.~(\ref{eq:upstair_vel_reward}) with the coefficients $\gamma_{vel}=1$, and $v^*=0.5~m/s$ for normal, soldier styles while $v^*=0.3~m/s$ for hopping style.
To further generate smoother motions, we use torque minimizing reward similar to Eq.~(\ref{eq:torque_reward}) for training normal and soldier style walking examples.
\paragraph{Cart Pulling}
In this scenario, the task is to pull a cart to the target position in the distance of $2~m$.
Therefore, we switch character root to the cart base in the formulation of Eq.~(\ref{eq:upstair_pos_reward}, \ref{eq:upstair_vel_reward}, \ref{eq:upstair_total_reward}) to shape cart tracking reward $r_{cart}$ where coefficients are $v^*=0.5~m/s,~\gamma_{pos} = 0.5, ~\gamma_{vel} = 64,~w_{pos} = 0.8, ~w_{vel}= 0.2$.
To properly guide both hands to play a key role in pulling the cart, we introduce hand reaching reward $r_{hand}$ to place hands on the proper area for the firm grasps.
$r_{hand}$ not only induces hand approaching to targets but also serves as a constraint. 
Specifically, by minimizing the search space of the positions of both hands, we can  better coordinate the learned grasping and the reference mimicking body motions.
To this end, we multiply a binary indicator $c$ for valid reaching, which in this example, examines whether a hand is located above the handle or not.
Accordingly, the reward functions can be represented as
\begin{equation}
    \label{eq:cart_hand_reward}
    r_{hand} = \prod_{n\in\{r, l\}} c_n\cdot\exp[-\gamma_{hand}\cdot \|p_n\|^2],
\end{equation}
\begin{equation}
    \label{eq:cart_total_reward}
    r^g = r_{hand}\cdot(w_{cart}\cdot r_{cart} + w_{hand})
\end{equation}
where $\{r, l\}$ indicates right and left hand, $p_n$ is the displacement from a hand to the target hand position, and $\gamma_{hand}=10,~w_{cart}=0.8,~w_{hand}=0.2$.

\paragraph{Bar Hanging}
In this example, the agent jumps to hang on the $2~m$ high horizontal bar, and subsequently hops to the left or right $60~cm$ along the bar for every 2 seconds.
To enable the transition to the hopping from the stable hanging pose, we use curriculum learning where the hand targets are fixed in the position so that no hopping is induced in the earlier phase of the training.
We terminate an episode if one of two feet remains in contact with the ground after $0.7 s$ from the state initialization.
Task reward is designed in a similar way to Eq.~(\ref{eq:cart_hand_reward}) with the parameter $\gamma_{hand} = 3$.
In this example, the reaching indicator $c$ checks if each palm is sufficiently aligned with the front direction of a character.
In the absence of the reaching indicator, the agent easily gets stuck in the local minima that yields unnatural poses of both hands as learning the transition between jumping and hanging motions is challenging.
\paragraph{Barbell Lifting}
Similar to \textit{Cart Pulling}, the objective in this task is to move a physical object to a desired pose.
Therefore, we modify task reward function in Eq.~(\ref{eq:cart_total_reward}) by replacing a cart tracking reward $r_{cart}$ with barbell tracking reward $r_{bbl}$ while maintaining the other values.
In contrast to previous interaction scenarios, we rather use real-numbered indicator $c$, which corresponds to the cosine value of the angle between the vector that heads the inner side of the palm and the vector that heads the barbell.
Two vectors here are alike the $d_h$ and $d_r$ respectively, in Figure~\ref{fig:supplementary_vector_description}.
Note indicator $c$ is scaled in $[0,~1]$ to maintain the task reward in a normalized scale.
Also, in order to enhance stability of the barbell lifting motions, we incorporate the barbell balance reward $r_{bal}$ described in the previous work~\cite{lee2022deep}.
As a result, the total reward is as follows:
\begin{equation}
    \label{eq:barbell_total_reward}
    r^g = r_{hand}\cdot(w_{bbl}\cdot r_{bbl} \cdot r_{bal} + w_{hand}),
\end{equation}
where $w_{bbl}=0.8$ and $w_{hand}=0.2$.

\paragraph{Rope Climbing}
In this example, each episode starts by throwing the agent toward the vertical rope composed with 30 units of $16~cm$-lengthed capsules.
Each of the left or right hand is set to grasp an upper unit capsule to procedurally generate the rope climbing motion.
We find that initializing the agent only with the right-hand up hanging leads to biased training and hampers the agent from learning hand alternation. 
To alleviate this problem, we randomize the initial target poses of both hands for each episode.
We formulate the task reward in a similar way to \textit{Bar Hanging} which can be represented as Eq.~(\ref{eq:cart_hand_reward}), but we slightly modify the formula as 
 the movement of both hands occur in alternating manner.
Specifically, we add a small offset $s$ to the reward of each hand so that the reward emitted by the other hand can be maintained when the target of one hand has just changed.
In summary, the finalized task reward is
\begin{equation}
    \label{eq:cart_hand_reward}
    r^g = \prod_{n\in\{r, l\}} c_n\cdot \{(1-s)\cdot\exp[-\gamma_{hand}\cdot \|p_n\|^2] + s\},
\end{equation}
where $\gamma_{hand}$ is 128 for the first hang, 16 for the rest of the climbing phase, and the offset $s$ is 0.05.
Note the hand pose indicator $c$ is calculated as a normalized cosine form similarly in \textit{Babell Lifting}.
\newpage
\subsection{Hyperparameters}\label{sec:hyperparameters}
\begin{table}[h]
  \caption{Values of hyperparamters for training.}
  \label{tab:comparison}
  \begin{tabular}{c|c}
    \toprule
    Parameters & value \\
    \midrule
    learning rate & 5e-5 \\
    $\gamma$ for GAE    & 0.99\\
    $\lambda$ for GAE    & 0.95\\
    clip range $\delta$ for PPO & 0.2 \\
    KL threshold for PPO & 0.008 \\
    batch size for PPO & 32768 \\
    batch size for PMP & 4096 \\
    demo buffer size for PMP & 2e5\\
    replay buffer size for PMP & 1e6 \\
    number of environments & 4096 \\
    horizon length & 16 \\
  \bottomrule
\end{tabular}
\end{table}

% % Figure - only
% \subsection{Locosight / UpstairCarry Frames}
% \subsection{Walking Analysis}
% \subsection{Interaction Scenes Snapshots}

\end{document}